\documentclass[11pt,a4paper]{article}

\usepackage{amssymb}
\usepackage{amsmath} 
\usepackage{oldgerm}
\usepackage{amsfonts}
\usepackage{euscript}
\usepackage[dvips]{epsfig}
\usepackage{here}

\begin{document}

\newcommand{\vr}{{\mathbf r}}
\newcommand{\vrt}{\tilde{\mathbf{r}}}
\newcommand{\vrp}{{\mathbf r'}}
\newcommand{\vy}{{\mathbf y }}
\newcommand{\vyp}{{\mathbf y}'}
\newcommand{\be}{\begin{equation}}
\newcommand{\en}{\end{equation}}
\newcommand{\ew}{\epsilon_{\rm w}}
\newcommand{\esolv}{\epsilon_{\rm s}}
\newcommand{\vw}{v_{\rm w}}
\newcommand{\hvw}{\hat{v}_{\rm w}}
\newcommand{\vb}{v_{\rm {\scriptscriptstyle B}}}
\newcommand{\vvw}{v_{\rm w} (\vr;\vrp)}
\newcommand{\dw}{\Delta_{\rm el}}
\newcommand{\rw}{\rho_{\rm w} (x)}
\newcommand{\kd}{\kappa_{\scriptscriptstyle D}}
\newcommand{\zb}{\overline{z}}
\newcommand{\zbu}{\overline{z}^{[1]}}
\newcommand{\zbd}{\overline{z}^{[2]}}
\newcommand{\zbj}{\overline{z}^{[{\rm j}]}}
\newcommand{\zbau}{\overline{z}_\alpha^{[1]}}
\newcommand{\zbad}{\overline{z}_\alpha^{[2]}}
\newcommand{\zbgu}{\overline{z}_\gamma^{[1]}}
\newcommand{\zbgd}{\overline{z}_\gamma^{[2]}}
\newcommand{\zbaj}{\overline{z}_\alpha^{[j]}}
\newcommand{\zbabu}{\overline{z}_{\alpha{\rm b}}^{[1]}}
\newcommand{\zbabd}{\overline{z}_{\alpha{\rm b}}^{[2]}}
\newcommand{\roaw}{\rho_{\alpha }(x;\beta,\{z_\gamma \})}
\newcommand{\roawn}{\rho_{\alpha  {\rm w}}}
\newcommand{\cp}{{\cal P}}
\newcommand{\cg}{{\mathbb{G}}}
\newcommand{\cpp}{{\mathbb{P}}}
\newcommand{\cs}{{\mathbb{S}}}
\newcommand{\fu}{\phi_1}
\newcommand{\fd}{\phi_2}
\newcommand{\usr}{U_{\scriptscriptstyle{\rm SR}}}
\newcommand{\Uelec}{U_{\rm elect}}
\newcommand{\vsr}{v_{\scriptscriptstyle{\rm SR}}}
\newcommand{\Vsr}{V_{\scriptscriptstyle{\rm SR}}}
\newcommand{\hVsr}{\hat{V}_{\scriptscriptstyle{\rm SR}}}
\newcommand{\Vself}{V_{\scriptscriptstyle{\rm self}}}
\newcommand{\hVself}{\hat{V}_{\scriptscriptstyle{\rm self}}}
\newcommand{\Vselfsc}{\Vself^{{\sssc}}}
\newcommand{\Vselfsce}{\Vself^{{\sssc} \, *}}
\newcommand{\row}{\rho_{\alpha \, {\rm w}}}
\newcommand{\roa}{\rho_{\alpha}}
\newcommand{\roab}{\rho_{\alpha}^{\scriptscriptstyle {\rm B}}}
\newcommand{\fcc}{f^{\rm c \, c}}
\newcommand{\Fcc}{F^{\rm c \, c}}
\newcommand{\Fcm}{F^{\rm c \, m}}
\newcommand{\Fmc}{F^{\rm m \, c}}
\newcommand{\Fmm}{F^{\rm m \, m}}
\newcommand{\ft}{f_{\scriptscriptstyle{\rm T}}}
\newcommand{\Frt}{F_{\scriptscriptstyle{\rm RT}}}
\newcommand{\Ir}{I_{\rm r}}
\newcommand{\rr}{{\EuScript{R}}}
\newcommand{\kud}{\kappa_1^2}
\newcommand{\kdd}{\kappa_2^2}
\newcommand{\kudb}{{\overline{\kappa_1}}^2}
\newcommand{\kddb}{{\overline{\kappa_2}}^2}
\newcommand{\ead}{e_\alpha^2}
\newcommand{\xt}{\tilde{x}}
\newcommand{\vq}{{\bf q}}
\newcommand{\kz}{\kappa_{\rm z}}
\newcommand{\fj}{\phi_{{\rm j}}}
\newcommand{\phit}{\widetilde{\phi}}
\newcommand{\fjt}{\tilde{\phi}_{\rm j}}
\newcommand{\fut}{\tilde{\phi}_{1}}
\newcommand{\fjtp}{\tilde{\phi}_{\rm j}^{*}}
\newcommand{\kzj}{\kappa_{j}}
\newcommand{\kzu}{\kappa_{1}}
\newcommand{\kzd}{\kappa_{2}}
\newcommand{\fub}{\phi_{1{\rm b}}}
\newcommand{\xxq}{(\xt,\xt', \vq)}
\newcommand{\eq}{E_{\rm q}}
\newcommand{\uj}{U_{\rm j}}
\newcommand{\Uj}{U_{\rm j}}
\newcommand{\hjp}{h_{{\rm j} }^{+*}}
\newcommand{\hjps}{h_{{\rm j} }^{+}}
\newcommand{\hjm}{h_{{\rm j} }^{-*}}
\newcommand{\hjms}{h_{{\rm j} }^{-}}
\newcommand{\bjt}{\tilde{b}_{{\rm j}}}
\newcommand{\bt}{\tilde{b}}
\newcommand{\rac}{\sqrt{1+q^2}}
\newcommand{\racm}{\sqrt{t^2-1}}
\newcommand{\Hjp}{H_{{\rm j}}^{+}}
\newcommand{\Hjm}{H_{{\rm j}}^{ -}}
\newcommand{\Hjme}{H_{{\rm j}}^{ - *}}
\newcommand{\eps}{\varepsilon}
\newcommand{\lt}{\tilde{l}}
\newcommand{\Ee}{\underset{\eps \rightarrow 0}{\rm Exp}}
\newcommand{\El}{\underset{l \rightarrow 0}{\rm Exp}}
\newcommand{\Elt}{\underset{\lt \rightarrow 0}{\rm Exp}}
\newcommand{\Eltseinf}{\underset{ \lt/\eps \rightarrow  \infty}{\rm Exp}}
\newcommand{\Eltinf}{\underset{ \lt \rightarrow  \infty}{\rm Exp}}
\newcommand{\Eu}{\underset{u \rightarrow 0}{\rm Exp}}
\newcommand{\Ujl}{U_{{\rm j}}^{{\rm lin}}}
\newcommand{\Ujlin}{U_{{\rm j}}^{{\rm lin}}}
\newcommand{\Od}{{\cal O}_{{\rm exp}} (\eps^2)}
\newcommand{\fz}{\phi(\dw=0)}
\newcommand{\fez}{\phi^{(0)}}
\newcommand{\ftz}{\widetilde{\phi}^{(0)}}
\newcommand{\fbz}{\phi_{\rm b}}
\newcommand{\ftb}{\widetilde{\phi}_{\scriptscriptstyle \rm B}}
\newcommand{\hpz}{h_{\scriptscriptstyle \rm HW}^{+*}}
\newcommand{\Hjpm}{H_{{\rm j}}^{\pm}}
\newcommand{\Hupm}{H_{{\rm 1}}^{\pm}}
\newcommand{\fuz}{\phi_1^{(0)}}
\newcommand{\epsa}{\varepsilon_\alpha}
\newcommand{\epsg}{\varepsilon_\gamma}
\newcommand{\Oda}{{\cal O}_{{\rm exp}} (\epsa^2)}
\newcommand{\Lv}{L(\kzu x; \kzu b ; \dw)}
\newcommand{\Lvd}{L(\kzd x; \kzd b ; \dw)}
\newcommand{\Zv}{Z( \vq; \bt , \dw)}
\newcommand{\hpzv}{h_{\scriptscriptstyle \rm HW}^{+} }
\newcommand{\htpzv}{\tilde{h}_{\scriptscriptstyle \rm HW}^{+} }
\newcommand{\htphw}{\tilde{h}_{\scriptscriptstyle \rm HW}^{+} }
\newcommand{\Lz}{L}
\newcommand{\Lt}{L^{\scriptscriptstyle T}}
\newcommand{\Lhw}{L^{\scriptscriptstyle HW}}
\newcommand{\Oua}{{\mathcal O}\left( \epsa, \frac{\beta \ead}{b}\right)}
\newcommand{\oua}{{\mathcal o}\left( \epsa, \frac{\beta \ead}{b}\right)}
\newcommand{\fb}{\phi_{{\rm b}}}
\newcommand{\fuu}{\phi_1^{(1)}}
\newcommand{\Oeps}{{\mathcal O} (\eps)}
\newcommand{\Oexpeps}{{\mathcal O}_{{\rm exp}} (\epsd)}
\newcommand{\Oexp}{{\mathcal O}_{{\rm exp}}}
\newcommand{\roawx}{\rho_{\alpha {\rm w}}(x)}
\newcommand{\cO}{{\cal O}}
\newcommand{\fdt}{\tilde{\phi}_2}
\newcommand{\vt}{{\bf t }}
\newcommand{\Frtt}{{\tilde F}_{\scriptscriptstyle{\rm RT}}}
\newcommand{\epsaap}{\varepsilon_{\alpha\gamma}}
\newcommand{\vu}{{\mathbf u}}
\newcommand{\wzpg}{w_0 (\xt '; \eps_\gamma, \dw)}
\newcommand{\wza}{w_0 (\xt ; \eps_\alpha, \dw)}
\newcommand{\ku}{\kappa_1}
\newcommand{\df}{\delta \phi}
\newcommand{\Ht}{\tilde{H}}
\newcommand{\OP}{\mathcal{O P}}
\newcommand{\OPU}{\mathcal{O P}_{{\rm U}}}
\newcommand{\OPUU}{\mathcal{O P}_{{\rm U}} \{ 1 \}(x) }
\newcommand{\OPUUn}{\mathcal{O P}_{{\rm U}} \{ 1 \} }
\newcommand{\OPUlinU}{\mathcal{O P}_{{\rm U_{\rm{lin}}}} \{ 1 \}(x)}
\newcommand{\OPUlinUn}{\mathcal{O P}_{{\rm U_{\rm{lin}}}} \{ 1 \}}
\newcommand{\OPUf}{\mathcal{O P}_{{\rm U}} \{ f \}(x) }
\newcommand{\ech}{\underset{\rightharpoondown}{\leftharpoonup} }
\newcommand{\Ei}{{\rm{Ei}}}
\newcommand{\Oc}{\mathcal{O}}
\newcommand{\psO}{\tilde{\mathcal{O}}}
\newcommand{\psOd}{\tilde{\mathcal{O}}\left( \eps^2 \right)}
\newcommand{\C}{\mathbf{C}}
\newcommand{\liml}{\underset{\eps \ll l \ll 1}{Lim}}
\newcommand{\Ulin}{U_{\rm{lin}}}
\newcommand{\Uuelin}{U_1^{* \, {\rm lin}}}
\newcommand{\Udelin}{U_2^{* \, {\rm lin}}}
\newcommand{\Uet}{\left( e^{\frac{\eps}{t}}-1 \right)}
\newcommand{\Uev}{\left( e^{\frac{\eps}{v}}-1 \right)}
\newcommand{\Lb}{\bar{L}}
\newcommand{\Ma}{M_\alpha}
\newcommand{\Mb}{\bar{M}}
\newcommand{\Mba}{{\bar{M}}_\alpha}
\newcommand{\epsd}{\eps_{{\scriptscriptstyle D}}}
\newcommand{\race}{\sqrt{|\eq|- \eta }\, \,  }
\newcommand{\hthw}{\tilde{h}_{{\scriptscriptstyle H \, W}}^{+}}
\newcommand{\Hjpe}{H_{{\rm j}}^{ + \, *}}
\newcommand{\hjep}{h_{{\rm j}}^{ + \, *}}
\newcommand{\Ujbmax}{{\bar U}_{\rm j}^{\rm max}}
\newcommand{\LUj}{ {\cal L}_{{\scriptscriptstyle U}_{\rm j }}}
\newcommand{\LUjlin}{ {\cal L}_{{\scriptscriptstyle U}_{\rm j
      }^{{\rm lin}}}}
\newcommand{\LUjbmax}{ {\cal L}_{{\scriptscriptstyle V}_{\rm
      j}^{\rm max}}}
\newcommand{\LmUjmax}{ {\cal L}_{-{\scriptscriptstyle U }_{\rm
      j}^{{\rm max}}}}
\newcommand{\LdifU}{ {\cal L}_{{\scriptscriptstyle U }_{\scriptscriptstyle
      j} - {\scriptscriptstyle U }_{\scriptscriptstyle
      j}^{{\rm  lin}}}}
\newcommand{\Lust}{ {\cal L}_{1/t}}
\newcommand{\Lgj}{ {\cal L}_{g_j}}
\newcommand{\HUj}{H_{{\scriptscriptstyle U}_{\rm j}}}
\newcommand{\HUjlin}{H_{{\scriptscriptstyle U}_{\rm j}^{{\rm lin}}}}
\newcommand{\RUj}{R_{{\scriptscriptstyle U}_{\rm j}}}
\newcommand{\RUjlin}{R_{{\scriptscriptstyle U}_{\rm j}^{{\rm lin}}}}
\newcommand{\HmUjmax}{H_{- {\scriptscriptstyle U}_{\rm j}^{{\rm max}}}}
\newcommand{\Uue}{U_1^*}
\newcommand{\Ude}{U_2^*}
\newcommand{\LU}{{\cal L}_{{\scriptscriptstyle U}}}
\newcommand{\LabsU}{{\cal L}_{\left| {\scriptscriptstyle U}\right|}}
\newcommand{\Labsg}{{\cal L}_{g}}
\newcommand{\Lmepsabsg}{{\cal L}_{- \eps g}}
\newcommand{\nth}{\mbox{n}^{\mbox{th}}}
\newcommand{\HU}{H_{{\scriptscriptstyle U}}}
\newcommand{\Hepsst}{H_{\eps / t}}
\newcommand{\Wepssdusd}{W_{\eps/2 \, , \, 1/2}}
\newcommand{\Uumax}{U_1^{{\rm max}}}
\newcommand{\Udmax}{U_2^{{\rm max}}}
\newcommand{\Ujmax}{U_{{\rm j}}^{{\rm max}}}
\newcommand{\Fj}{F_{{\rm j}}}
\newcommand{\epsj}{\eps_{{\rm j}}}
\newcommand{\uua}{u_{1 \, \alpha}}
\newcommand{\uda}{u_{2 \, \alpha}}
\newcommand{\Uulin}{U_1^{{\rm lin}}}
\newcommand{\hupe}{h_1^{+ *}}
\newcommand{\hume}{h_1^{- *}}
\newcommand{\hupme}{h_1^{\pm *}}
\newcommand{\hdpe}{h_2^{+ *}}
\newcommand{\Udlin}{U_2^{{\rm lin}}}
\newcommand{\roax}{\rho_\alpha (x)}
\newcommand{\eg}{e_\gamma}
\newcommand{\rog}{\rho_\gamma}
\newcommand{\hagrrp}{h_{\alpha \gamma} (\vr; \vr')}
\newcommand{\fcm}{\phi^{{\scriptscriptstyle MF}}}
\newcommand{\fcmlin}{\phi^{{\scriptscriptstyle MF, \,}{\rm lin}}_{z}}
\newcommand{\roacm}{\rho_\alpha^{{\scriptscriptstyle MF}}}
\newcommand{\ea}{e_\alpha}
\newcommand{\eap}{e_{\alpha'}}
\newcommand{\Vtotcm}{V_{{\rm tot}}^{{\scriptscriptstyle MF}}}
\newcommand{\dVtotcm}{\delta V_{{\rm tot}}^{{\scriptscriptstyle MF}}}
\newcommand{\zacm}{z_\alpha^{{\scriptscriptstyle MF}}}
\newcommand{\zgcm}{z_\gamma^{{\scriptscriptstyle MF}}}
\newcommand{\rocmlin}{\rho_\alpha^{{\scriptscriptstyle MF, \,}{\rm lin}}}
\newcommand{\Vtotcmlin}{V_{{\rm tot}}^{{\scriptscriptstyle MF, \,}{\rm
      lin}}}
\newcommand{\dVtotcmlin}{\delta V_{{\rm tot}}^{{\scriptscriptstyle MF, \,}{\rm
      lin}}}
\newcommand{\egd}{e_\gamma^2}
\newcommand{\Vtot}{V_{{\rm tot}}}
\newcommand{\Pb}{P^{{\rm \scriptscriptstyle B}}}
\newcommand{\kdb}{k_{{\scriptscriptstyle D}}}
\newcommand{\kb}{k_{{\scriptscriptstyle B}}}
\newcommand{\roahw}{\rho_\alpha^{{\scriptscriptstyle HW}}}
\newcommand{\Mbhw}{\bar{M}^{{\scriptscriptstyle HW}}}
\newcommand{\Lbhw}{\bar{L}^{{\scriptscriptstyle HW}}}
\newcommand{\Mhw}{M^{{\scriptscriptstyle HW}}}
\newcommand{\rogp}{\rho_{\gamma'}}
\newcommand{\egp}{e_{\gamma'}}
\newcommand{\Vtothw}{\Vtot^{{\scriptscriptstyle HW}}}
\newcommand{\rogb}{\rho_{\gamma}^ {\scriptscriptstyle{\rm B}}}
\newcommand{\Vas}{V_{{\rm as}}}
\newcommand{\Pas}{\Phi_{{\rm as}}}
\newcommand{\roaasym}{\rho_\alpha^{{ \rm asym}}}
\newcommand{\roasym}{\rho_\alpha^{{ \rm sym}}}
\newcommand{\equivx}{\underset{x \rightarrow + \infty}{\sim}}
\newcommand{\soma}{\sum_\alpha}
\newcommand{\rohw}{\rho^{{\scriptscriptstyle HW}}}
\newcommand{\Phw}{\Phi^{{\scriptscriptstyle HW}}}
\newcommand{\ropb}{\rho_{+}^ {\scriptscriptstyle{\rm B}}}
\newcommand{\sscm}{{\scriptscriptstyle MF}}
\newcommand{\Pcm}{\Phi^{\sscm}}
\newcommand{\Pcmlin}{\Phi^{\sscm, {\rm lin}}}
\newcommand{\rop}{\rho_+}
\newcommand{\rom}{\rho_-}
\newcommand{\romb}{\rho_{-}^ {{\rm B}}}
\newcommand{\kbe}{\kd \beta e^2}
\newcommand{\kdbf}{\kd b}\newcommand{\ei}{e_{{\rm i}}}
\newcommand{\ej}{e_{{\rm j}}}
\newcommand{\rj}{{\vr}_{{\rm j}}}
\newcommand{\ri}{{\vr}_{{\rm i}}}
\newcommand{\vc}{v_{{\rm c}}}
\newcommand{\vux}{{\mathbf u}_x}
\newcommand{\ssB}{{\scriptscriptstyle B}}
\newcommand{\roainf}{\rho_{\alpha}^{\ssB}}
\newcommand{\roarmu}{\rho_\alpha \left( \vr ; \{ \mu_\gamma \} \right)}
\newcommand{\QdD}{Q_{\delta \calD}}
\newcommand{\norme}{\vert {\mathbf e} \vert}
\newcommand{\vmu}{{\boldsymbol{\mu}}}
\newcommand{\vN}{{\mathbf N}}
\newcommand{\ve}{{\mathbf e}}
\newcommand{\cC}{\hat{C}}
\newcommand{\roaz}{\rho_\alpha (0)}
\newcommand{\emusepu}{\left( \frac{\ew - 1}{\ew + 1} \right)}
\newcommand{\emusepup}{\left( \frac{\varepsilon_w - 1}{\varepsilon_w +
      1} \right)}
\newcommand{\umesupe}{\left( \frac{1-\ew}{1+\ew} \right)}
\newcommand{\umesupep}{\left( \frac{1-\varepsilon_w}{1+\varepsilon_w}
  \right)}
\newcommand{\vsdu}{v_{{\scriptscriptstyle SD}}}
\newcommand{\vsdd}{V_{{\scriptscriptstyle SD}}}
\newcommand{\zux}{z^{[1]} (x)}
\newcommand{\zaux}{z_{\alpha}^{[1]} (x)}
\newcommand{\kudx}{\kappa_1^2 (x)}
\newcommand{\vl}{{\mathbf l}}
\newcommand{\kudinf}{(\kappa_1^{\ssB})^2 }
\newcommand{\kddx}{\kappa_2^2(x)}
\newcommand{\emur}{\epsilon_{{\scriptscriptstyle mur}}}
\newcommand{\Wax}{W_\alpha (x)}
\newcommand{\xid}{\xi_{{\scriptscriptstyle D}}}
\newcommand{\Gtd}{\Gamma^{3/2}}
\newcommand{\roinf}{\rho^{\ssB}}
\newcommand{\hA}{h_{{\scriptscriptstyle A}}}
\newcommand{\fireiriv}{v (\vr ; \{ e_i, \vr_i \}, V) } 
\newcommand{\Utelect}{U_{{\rm elect}}^*}
\newcommand{\muta}{\mu_\alpha^*}
\newcommand{\vp}{{\mathbf p}}
\newcommand{\laa}{\lambda_\alpha}
\newcommand{\laap}{\lambda_{\alpha'}}
\newcommand{\zta}{z_\alpha}
\newcommand{\ztg}{z_\gamma}
\newcommand{\cPi}{{\mathcal{P}}_i}
\newcommand{\somg}{\sum_\gamma}
\newcommand{\za}{z_\alpha}
\newcommand{\zg}{z_\gamma}
\newcommand{\vz}{{\mathbf 0}}
\newcommand{\Nb}{\bar{N}}
\newcommand{\Vsd}{\frac{V}{2}}
\newcommand{\bedsb}{\beta e^2 / b}
\newcommand{\sgegtrgb}{\somg \eg^3 \rogb}
\newcommand{\sgegdrgb}{\somg \eg^2 \rogb}

\newcommand{\pa}{p_\alpha}
\newcommand{\pg}{p_\gamma}
\newcommand{\diagFcc}{\mbox{\begin{picture}(58,25)(13,-3)
\put(20,0){\circle{10}}
\put(25,0){\line(1,0){40}}
\put(65,0){\circle*{5}}
\put(40,10){$\Fcc$}
\put(18,10){$\scriptstyle{x}$}
\put(63,10){$\scriptstyle x'$}
\end{picture}}}
\newcommand{\diagFccq}{\mbox{\begin{picture}(58,25)(13,-3)
\put(20,0){\circle{10}}
\put(25,0){\line(1,0){40}}
\put(65,0){\circle*{5}}
\put(40,10){$\Fcc$}
\end{picture}}}
\newcommand{\diagFrt}{\mbox{\begin{picture}(58,25)(13,-3)
\put(20,0){\circle{10}}
\put(25,0){\line(1,0){40}}
\put(65,0){\circle*{5}}
\put(40,10){$\Frt$}
\end{picture}}}
\newcommand{\diagFmc}{\mbox{\begin{picture}(58,25)(13,-3)
\put(20,0){\circle{10}}
\put(65,0){\vector(-1,0){40}}
\put(65,0){\circle*{5}}
\put(40,10){$\Fmc$}
\end{picture}}}
\newcommand{\diagFcm}{\mbox{\begin{picture}(58,25)(13,-3)
\put(20,0){\circle{10}}
\put(25,0){\vector(1,0){38}}
\put(65,0){\circle*{5}}
\put(40,10){$\Fcm$}
\end{picture}}}
\newcommand{\diagfcc}{\mbox{\begin{picture}(65,25)(13,-3)
\put(20,0){\circle{10}}
\put(25,0){\line(1,0){40}}
\put(70,0){\circle{10}}
\put(40,10){$\fcc$}
\end{picture}}}
\newcommand{\diagfcm}{\mbox{\begin{picture}(65,25)(13,-3)
\put(20,0){\circle{10}}
\put(25,0){\vector(1,0){40}}
\put(70,0){\circle{10}}
\put(40,10){$\fcmq$}
\end{picture}}}
\newcommand{\diagfmc}{\mbox{\begin{picture}(65,25)(13,-3)
\put(20,0){\circle{10}}
\put(65,0){\vector(-1,0){40}}
\put(70,0){\circle{10}}
\put(40,10){$\fmcq$}
\end{picture}}}
\newcommand{\diagfmm}{\mbox{\begin{picture}(65,25)(13,-3)
\put(20,0){\circle{10}}
\put(65,0){\vector(-1,0){40}}
\put(25,0){\vector(1,0){40}}
\put(70,0){\circle{10}}
\put(40,10){$\fmm$}
\end{picture}}}
\newcommand{\diagdfcc}{\mbox{\begin{picture}(70,25)(13,-3)
\put(20,0){\circle{10}}
\put(25,0){\line(1,0){45}}
\put(50,0){\circle*{5}}
\put(75,0){\circle{10}}
\put(60,10){$\fcc$}
\put(30,10){$\fcc$}
\put(50,-15){$\zb$}
\end{picture}}}
\newcommand{\diagtfcc}{\mbox{\begin{picture}(70,25)(13,-3)
\put(20,0){\circle{10}}
\put(25,0){\line(1,0){45}}
\put(39,0){\circle*{5}}
\put(57,0){\circle*{5}}
\put(75,0){\circle{10}}
\put(27,10){$\fcc$}
\put(43,10){$\fcc$}
\put(60,10){$\fcc$}
\put(36,-15){$\zb$}
\put(56,-15){$\zb$}
\end{picture}}}
\newcommand{\zat}{z_\alpha}
\newcommand{\Phib}{\Phi^{{\rm B}}}
\newcommand{\vecz}{{\mathbf 0}}
\newcommand{\so}{\sigma^{{\rm o}}}
\newcommand{\sd}{\sigma^{{\rm d}}}
\newcommand{\rpbsestd}{\frac{\sqrt{\pi} \beta^{3/2}}{\esolv^{3/2}}}
\newcommand{\setrsredr}{\frac{\somg \eg^3 \rogb}{\left( \sum_\delta 
e_\delta^2 \rho_{\delta}^{{\rm B}} \right)^{1/2}}}
\newcommand{\dsetrsredr}{\somg \frac{ \eg^3 \rogb}{\left( \sum_\delta 
e_\delta^2 \rho_{\delta}^{{\rm B}} \right)^{1/2}}}
\newcommand{\sssc}{{\scriptscriptstyle sc}}
\newcommand{\Vselfcm}{\Vself^{{\sssc}}}
\newcommand{\Vselfcmlin}{\Vself^{{\sssc}, {\rm lin}}}
\newcommand{\rocp}{\rho_{{\scriptscriptstyle OCP}}}
\newcommand{\Mocp}{M_{{\scriptscriptstyle OCP}}}
\newcommand{\ssocp}{{\scriptscriptstyle OCP}}
\newcommand{\Sbulk}{S^{\ssB}}
\newcommand{\rob}{\rho^{\ssB}}
\newcommand{\vvp}{{\mathbf p}}
\newcommand{\ssW}{{\scriptscriptstyle W}}
\newcommand{\ewz}{\varepsilon_{\ssW_0}}
\newcommand{\vF}{{\mathbf F}}
\newcommand{\hH}{\hat{H}}
\newcommand{\hT}{\hat{T}}
\newcommand{\hV}{\hat{V}}
\newcommand{\hU}{\hat{U}}
\newcommand{\hvp}{\hat{\vp}}
\newcommand{\krrp}{\boldsymbol{|}\vr, \vr' \boldsymbol{>}}
\newcommand{\TrL}{{\rm Tr}_\Lambda}
\newcommand{\mua}{\mu_\alpha}
\newcommand{\Xig}{\boldsymbol{\Xi}}
\newcommand{\nap}{n_p^{\alpha}}
\newcommand{\etaa}{\eta_\alpha}
\newcommand{\roag}{\rho_{\alpha \, \gamma}^{(2)}}
\newcommand{\ssT}{{\scriptscriptstyle T}}
\newcommand{\roagT}{\rho_{\alpha \, \gamma}^{(2) \, \ssT}}
\newcommand{\hag}{h_{\alpha \, \gamma}}
\newcommand{\brarp}{\boldsymbol{<} \vr' \boldsymbol{|}}
\newcommand{\brar}{\boldsymbol{<} \vr \boldsymbol{|}}
\newcommand{\brarurd}{\boldsymbol{<} \vr_1, \vr_2 \boldsymbol{|}}
\newcommand{\brardru}{\boldsymbol{<} \vr_2, \vr_1 \boldsymbol{|}}
\newcommand{\braxp}{\boldsymbol{<} x' \boldsymbol{|}}
\newcommand{\ketr}{\boldsymbol{|} \vr \boldsymbol{>}}
\newcommand{\ketrurd}{\boldsymbol{|} \vr_1, \vr_2 \boldsymbol{>}}
\newcommand{\ketx}{\boldsymbol{|} x \boldsymbol{>}}
\newcommand{\hAop}{\hat{A}}
\newcommand{\hB}{\hat{B}}
\newcommand{\hUop}{\hat{U}}
\newcommand{\vxi}{\boldsymbol{\xi}}
\newcommand{\xix}{\xi_{{\rm x}}}
\newcommand{\xip}{\vxi_{\parallel}}
\newcommand{\md}{\frak{D}}
\newcommand{\Dx}{\frak{D}_{{\rm x}}}
\newcommand{\D}{\frak{D}}
\newcommand{\Erfc}{{\rm Erfc}}
\newcommand{\lsat}{\left( \lambda / a \right)^3}
\newcommand{\Lr}{\frak{L}}
\newcommand{\vwmvb}{\left[ \vw - \vb \right]}
\newcommand{\arxi}{\left( \alpha, \vr, \vxi \right)}
\newcommand{\rol}{\rho ( \Lr )}
\newcommand{\Lra}{\frak{L}_{{\rm a}}}
\newcommand{\Lrb}{\frak{L}_{{\rm b}}}
\newcommand{\vcc}{v^{{\rm c \, c }}}
\newcommand{\vcm}{v^{{\rm c \, m }}}
\newcommand{\vmc}{v^{{\rm m \, c }}}
\newcommand{\vmm}{v^{{\rm m \, m }}}
\newcommand{\bij}{\beta_{ i \, j}}
\newcommand{\zbl}{\zb ( \Lr )}
\newcommand{\fcmq}{f^{{\rm c \, m}}}
\newcommand{\fmcq}{f^{{\rm m \, c}}}
\newcommand{\fmm}{f^{{\rm m \, m}}}
\newcommand{\Lri}{\Lr_i}
\newcommand{\Lrj}{\Lr_j}
\newcommand{\Lrz}{\Lr_0}
\newcommand{\Lru}{\Lr_1}
\newcommand{\Lrd}{\Lr_2}
\newcommand{\ftt}{f_{\ssT \ssT}}
\newcommand{\Arz}{\frak{A}_0}
\newcommand{\fnc}{f^{- \, {\rm c}}}
\newcommand{\fcn}{f^{{\rm c} \, - }}
\newcommand{\Fnc}{F^{- \, {\rm c}}}
\newcommand{\Fcn}{F^{{\rm c} \, - }}
\newcommand{\kub}{\bar{\kappa}_1}
\newcommand{\dxdslad}{\frac{2 x^2}{\laa^2}}
\newcommand{\ssQ}{{\scriptscriptstyle Q}}
\newcommand{\ZQ}{Z^{\ssQ}}
\newcommand{\Lbq}{\bar{L}^{\ssQ}}
\newcommand{\kdl}{\kzd \lambda}
\newcommand{\lag}{\lambda_\gamma}
\newcommand{\lau}{\lambda_1}
\newcommand{\izuds}{\int_0^1 ds \,}
\newcommand{\nabg}{\boldsymbol{\nabla}}
\newcommand{\Mbq}{\Mb^{\ssQ}}
\newcommand{\Br}{\frak{B}}
\newcommand{\vf}{{\mathcal U}}
\newcommand{\plsapd}{\left( \frac{\lambda}{a} \right)}
\newcommand{\plsap}{\left( \lambda / a \right)}
\newcommand{\lsa}{ \frac{\lambda}{a}}
\newcommand{\kdebl}{\kd \lambda}
\newcommand{\Psia}{\Psi_{\alpha}}
\newcommand{\ma}{m_{\alpha}}
\newcommand{\UW}{U_{\scriptscriptstyle W}}
\newcommand{\vcW}{v_{\scriptscriptstyle CW}}
\newcommand{\vSR}{v_{\scriptscriptstyle SR}}
\newcommand{\eW}{\varepsilon_{\scriptscriptstyle W}}
\newcommand{\vcb}{v_{{\scriptscriptstyle C}{\rm bulk}}}
\newcommand{\Dpnu}{D_{\scriptscriptstyle W}^0}
\newcommand{\Dp}{{\overline D_{\scriptscriptstyle W}^0}}
\newcommand{\Dpt}{{\overline D_{\scriptscriptstyle W}}}
\newcommand{\DptD}{{\overline D_{\scriptscriptstyle DW}}}
\newcommand{\xiD}{\xi_{\scriptscriptstyle B}}
\newcommand{\alp}{\scriptscriptstyle\alpha}
\newcommand{\gam}{\scriptscriptstyle\gamma}
\newcommand{\rhoT}{\rho^{(2){\scriptscriptstyle T}}}
\newcommand{\phiW}{\phi_{\scriptscriptstyle W}}
\newcommand{\phiDW}{\phi_{\scriptscriptstyle DW}}
\newcommand{\phiDb}{\phi_{{\scriptscriptstyle D}{\rm bulk}}}
\newcommand{\kp}{{\bf k}_{\scriptscriptstyle //}}
\newcommand{\x}{{\widetilde x}}
\newcommand{\g}{{\scriptscriptstyle \gamma}}
\newcommand{{\hclW}}{h_{{\rm cl},{\scriptscriptstyle W}}}
\newcommand{{\hclDW}}{h_{{\rm cl},{\scriptscriptstyle DW}}}
\newcommand{\phiMFW}{\phi_{\scriptscriptstyle MFW}}
\newcommand{{\hclMFW}}{h_{{\rm cl},{\scriptscriptstyle MFW}}}
\newcommand{\rhodT}{\rho^{\scriptscriptstyle(2) T}}
\newcommand{\rhoW}{\rho_{\scriptscriptstyle W}}
\newcommand{\hW}{h_{\scriptscriptstyle W}}
\newcommand{\bzero}{\boldsymbol{0}}
\newcommand{\bnab}{\boldsymbol{\nabla}}
\newcommand{\bxi}{\boldsymbol{\xi}}
\def\a{{\scriptscriptstyle\alpha}}
\def\g{{\scriptscriptstyle\gamma}}
\def\d{{\scriptscriptstyle\delta}}
\def\y{\vy}
\newcommand{\vyt}{\tilde{\mathbf{y}}}
\newcommand{\felect}{f_{\rm elect}}
\newcommand{\fw}{f_{\omega}}
\newcommand{\fcl}{f_{\rm cl}}
\newcommand{\ssdd}{{\scriptscriptstyle 2D}}
\newcommand{\Gdd}{\Gamma_{\ssdd}}
\newcommand{\sun}{\sigma_1}
\newcommand{\sdeux}{\sigma_2}
\newcommand{\Ener}{\mathcal{E}}
\newcommand{\Erf}{{\rm Erf}}
\newcommand{\cnuag}{c_{{\rm nuage}}}
\newcommand{\Oetad}{\cO (\eta^2)}
\newcommand{\RLj}{R_{L_{{\rm j}}}}
\newcommand{\benote}{\begin{dinglist}{56}}
\newcommand{\ennote}{\end{dinglist}}
\newcommand{\calD}{{\mathcal D}}
\newcommand{\bgy}{{\textsc{bgy}} }
\newcommand{\ns}{n_{{\rm s}}}

\thispagestyle{empty}

\title{\bf Density profiles in a classical Coulomb fluid near a
  dielectric wall. I. Mean-field scheme}
\author{Jean-No\"el AQUA and Fran\c{c}oise CORNU
\\
Laboratoire de Physique Th\'eorique \thanks{Laboratoire associ\'e
au Centre National de la Recherche Scientifique - UMR 8627} \\
B\^atiment 210, Universit\'e Paris-Sud\\ 
91405 Orsay Cedex, France}

\maketitle
\begin{abstract}

The equilibrium density profiles in a classical multicomponent plas\-ma near a hard
wall made with a dielectric material characterized by a relative
dielectric constant $\ew$ are studied from the first Born-Green-Yvon (\bgy\!\!) equation 
combined with  Poisson equation in a regime where Coulomb coupling is weak
inside the fluid.  In order to prevent the collapse
between charges with opposite signs or between each charge and its
dielectric image inside the wall when $\ew >1$,  hard-core repulsions are
added to the Coulomb pair interaction.  The charge-image interaction cannot be 
treated perturbatively and the density profiles vary very fast in the vicinity of 
the wall when $\ew \neq 1$. The formal solution of the associated inhomogeneous
Debye-H\"uckel equations will be given in  Paper II, together with a systematic fugacity 
expansion which allows to retrieve the results obtained from the truncated \bgy hierarchy.
In the present paper 
the exact density profiles are calculated analytically up to first order
in the coupling parameter. The expressions show the interplay between
three effects~: the geometric repulsion from the impenetrable wall; the
electrostatic effective attraction ($\ew >1$) or repulsion ($\ew <1$)
due to its dielectric response; and the Coulomb  interaction between
each charge and the potential drop created by the electric layer which
appears as soon as the system is not  symmetric. We exhibit how
the charge density profile evolves between a structure with two
oppositely-charged layers and a three-layer organization when $\ew$
varies. (The case of two ideally conducting walls will be displayed elsewhere).  

{\bf KEYWORDS~:} Coulomb interactions, dielectric wall, \bgy equa\-tion,
inhomogeneous Debye-H\"uckel equation, electric layer.

\vskip 1cm 
\end{abstract}

\newpage

\numberwithin{equation}{section}

\section{Introduction}

The present paper provides new exact analytical perturbative results for the density
profiles of a classical Coulomb plasma in the vicinity of a polarizable
boundary. We consider a
multicomponent plasma, namely a system made of at
least two species of moving charges with opposite signs. The  linear 
electrostatic response  of the wall is described at a macroscopic 
level by a relative
dielectric constant $\ew$. ($\ew$ is the ratio of  the dielectric constants in 
the wall and in the half-space occupied by the Coulomb fluid).
The density profiles are obtained in a high-temperature 
(or low-density) limit which is realized for instance in an electrolyte 
solution. 

As shown in Paper II -- published just after the present paper -- this limit 
is the first-order 
result in a systematic expansion in powers of the Coulomb coupling 
parameter. This limit  can be retrieved from a mean-field approximation for the
first Born-Green-Yvon (\bgy\!) equation which leads to the resolution of
 inhomogeneous Debye-H\"uckel  equations. 
For the sake of pedagogy, the present paper  is devoted 
to the  mean-field interpretation, which should be more familiar to readers
interested in chemical physics, and to the discussion of the properties of the
electric layer. The exact derivation is postponed to Paper II
where we present two points: first, systematic resummations of Coulomb
divergencies in the framework of the 
grand-canonical ensemble; second, the resolution of the inhomogeneous
Debye-H\"uckel equations obeyed by the auxiliary effective potentials 
which arise from the latter resummations.

The Coulomb pair interaction $v(\vr;\vrp)$ between two unit charges located
respectively at $\vr$ and $\vrp$ near the dielectric wall, namely the solution of Poisson equation 
\begin{equation}
  \Delta_{\vrp} v(\vr;\vrp) = - 4 \pi \delta(\vr - \vrp)
  \label{pois}
\end{equation}
with the adequate electrostatic boundary conditions, reads
\begin{equation}
\vvw = \frac{1}{|\vr - \vrp|} - \dw \frac{1}{|\vr - \vrp ^*|}
\label{defvw}
\end{equation}
where $\dw \equiv (\ew-1)/(\ew+1) $ and $\vrp ^*$ is the image of
$\vrp$ with respect to the plane interface \cite{Jackson}.
When $\ew$ varies from $0$ to $+\infty$, $\dw$ ranges from $-1$ to
$1$. In a dielectric material $-1 < \dw < 1$. 
If the Coulomb fluid mimics an electrolyte in a solvent described
as a rigid continuum medium, $\ew$ is the relative
dielectric constant of the wall with respect to the solvent dielectric constant
$\esolv$ and the interaction potential $\vw$ in \eqref{defvw} is to be
multiplied by $1/\esolv$. The potential \eqref{defvw} may be seen as 
the sum of two contributions. The vacuum or ``bulk'' potential 
\begin{equation}
v_{\rm {\scriptscriptstyle B}} (\vr ; \vrp ) = \frac{1}{|\vr - \vrp | } 
\label{defvb}
\end{equation}
is the solution of 
\eqref{pois}
far away from any boundary or in the
vicinity of a wall with no electrostatic response ($\ew = 1 $). The
second term in \eqref{defvw} is the interaction with an ``image'' charge;
the latter describes the interaction with the polarization charge 
generated in the material by plasma charges. The corresponding self-energy 
of a charge $\ea$ at point $\vr$ -- namely the work needed to take one charge 
$\ea$ from the bulk to point $\vr$ in the vicinity of the wall -- is equal to $\ead \Vself (\vr)$ with
\begin{equation}
  \label{defVself}
\Vself (\vr) =  \frac{1}{2} \left[ \vw - \vb \right] (\vr, \vr)
\end{equation}
In \eqref{defVself} the factor $1/2$ in the interaction between a charge and its
image comes from the proportionality between the two charges.
In the following, the interface is 
perpendicular to the $x$-axis and located at $x\!=\!0$, and,
according to \eqref{defvw}, 
\begin{equation}
 \Vself (x) = -\dw \frac{1}{4 x}
\label{valueVself}
\end{equation}
When $\dw >0$ a hard-core repulsion from the wall must be introduced
in order to prevent the collapse of each charge with its image.  
For the sake of
simplicity, the range $b$ of the  repulsion from the wall is chosen to be the same
for all species in the present paper.
Even in
the bulk, a short-distance cut-off must be introduced in order to
prevent the collapse of the system due to the attraction between charges with
opposite signs. However this second cut-off proves not to arise in the densities
  at leading order in the Coulomb coupling parameter inside 
the fluid. (Indeed, in the first BGY hierarchy the variation of the 
density of every species 
 depends on correlations
only through an integral and the value of the latter integral at leading order
in the Coulomb coupling parameter is determined only by the behavior of correlations at
distances far larger than the short-distance cut-off.)

For a long time,  the short-distance singularity of
the charge-image interaction has prevented one from getting exact
results in the case $\ew \neq 1 $ at any distance from the wall for
either a generic multicomponent or a One-Component Plasma (OCP), namely 
a system made of only one moving charge species in a rigid 
neutralizing background. The self-consistent method 
introduced by Guernsey \cite{Guer70} for a plain wall ($\ew=1$)
was generalized for the first time to a
case where $\ew \neq 1$ by Alastuey \cite{Alas83}. This author dealt
with the OCP near a wall with a repulsive
electrostatic response ($\ew < 1 $) in the weak-coupling limit.
In this case the density vanishes on the wall and drastically varies
over the closest approach distance $\beta e^2$. The mean-field
 electrostatic potential $\Phi(x)$
created by the charge density profile 
is solution of an inhomogeneous Debye-H\"uckel equation where the inverse Debye
length depends on the distance from the wall and rapidly varies in its vicinity.
Alastuey solved the
equation for the mean-field value of $\Phi (x)$ and produced the corresponding profile 
density
but only for distances larger than the closest approach distance. (For
these distances a linearization may be performed and the equation for
$\Phi (x)$ is a  second-order linear differential
equation with constant parameters). The case $\ew >1$, where the
attractive response of the wall makes the density diverge exponentially
fast on the wall in the absence of any hard-core repulsion, was left
unsolved at any distance.

In Section 2 we introduce a self-consistent scheme for the determination of 
density
profiles in a multicomponent plasma from the first \bgy equation combined with
Poisson equation. By using the results of Paper II about 
the
solutions of the corresponding inhomogeneous Debye-H\"uckel equations,  we give 
their formal expressions {\it at any distance} from the wall
at first order in the coupling parameter  $\epsd $ inside the fluid, 
\begin{equation}
  \label{blibli}
  \epsd \equiv \frac{1}{2} \beta e^2 \kd
\end{equation}
In \eqref{blibli} $\beta$ is  the inverse temperature,
$\beta = 1 / \kb T$ where $\kb$ is Boltzmann 
constant,  $e$ is the typical charge in the plasma and 
$\kd$ is the inverse Debye length
\begin{equation}
  \label{defkdbis}
  \kd \equiv \sqrt{4 \pi \beta \soma \ead \roab}
\end{equation}
where $\roab$ is the bulk value of the density for species with index $\alpha$
 (and $\ea$ is the charge of species $\alpha$). The sum over $\alpha$ runs from 1 
to the number of species $\ns$. 
Every  density profile takes the form
\begin{equation}
  \label{tita}
  \roax = \roab \, \theta (x-b) \, e^{- \beta \ead \Vselfsc (x) } \left[ 1 -
  \beta \ea \Phi(x) \right]
\end{equation}
where $\Vselfsc (x)$ is a screened self-energy  and $\Phi(x)$ is the 
 electrostatic potential
created by the charge density profile $\sum_\gamma e_\gamma \rho_\gamma (x)$.
($\Phi(x)$ is set to $0$ in the bulk.)
 The Heaviside function 
$\theta(u)$ -- with
$\theta(u) = 1 $ if $u>0$ and $\theta(u) = 0 $ if $u<0$ -- describes the
geometric constraint enforced by the impenetrable wall.
The analytic expressions are obtained  in Section 3. $\Phi(x)$, given in 
 \eqref{vtotcmlin}, 
decays as $\exp (-\kd x)$ at large distances. 
$\Vselfsc (x)$ may be written as the sum 
\begin{equation}
  \label{undix}
  \Vselfsc (x) = \frac{\kd}{2} \Lb (\kd x; \kd b, \dw) - \frac{\dw}{4x}
  \exp(-2\kd x)
\end{equation}
$\Vselfsc (x)$ falls off as $\exp (-2 \kd x)/ 4x$ when $x$ goes to
infinity for all values of $\dw$. 
$(\kd/2)\Lb$ given in \eqref{valueL} and \eqref{decompL} 
arises mainly from the geometric deformation of the
screening cloud around a charge in the vicinity of the wall and remains finite
at any distance. On the contrary,
$-
\ead (\dw /4x) \exp(-2\kd x)$ is the part of the screened self-energy
originating from the bare self-energy $\ead\Vself$ \eqref{valueVself}
due to the dielectric response of the wall. The second term in $\Vselfsc (x)$ 
was derived for the first time in the case $\ew <1$ from a
phenomenological mean-field argument by Onsager and Samaras in 1934
\cite{Onsa&Sama34}. Its contribution to the density profile is crucial at
short distances. When $\ew < 1 $ ($\dw <0$) all charges are
electrostatically repelled by the wall, the short-distance repulsion
range $b$ can be set to zero and the profile density vanishes
exponentially fast at the contact ($x=0$) with the wall. On the contrary,
when $\ew > 1$ ($\dw >0$), all charges are attracted by the wall, $b$
must be kept finite and the contact value $\roa(b)$ increases as 
$\exp \left[ \dw \beta \ead \exp(- 2 \kd b) /(4b)\right] $ when $b$ becomes small.
In Section 3 we also derive the profile density in a OCP and we compare our result with 
that of  Ref. \cite{Alas83}. 

Section 4 is devoted to generic global properties of the plasma 
at the interface.
In Section 5  we study the case of a plain hard wall ($\ew =1$). The analytic expressions are rather 
simple and we can investigate the only two effects which interplay~: the geometric repulsion from the wall 
and the interaction with the electrostatic potential drop $\Phi (x)$ created by 
the electric layer itself. In the case of a symmetric two-component plasma, we
retrieve the results of \cite{Janco82I}. In Section
6 the generic properties of the density profiles when $\ew\neq 1$ are interpreted in terms of the competition between three 
effects~: the two ones already at stake in the vicinity of a plain hard wall plus the electrostatic (repulsive
or attractive) interaction due to the dielectric response of the wall. In particular, we exhibit how the structure of 
the charge density profile evolves from a double layer into a threefold layer and then into an inversed double
 layer when $\ew$ increases from the value $\ew = 1$.

%%%%%%%%%%%%%%%%%%%%%%%%%%%%%%%%%%%%%%%%%

\section{Self-consistent scheme in the weak-coupling re\-gi\-me}
\label{section2}

\subsection{ Exact first ${\rm { BGY}}$ equation}

The exact density profile $\roax$ is related to the Ursell function
$h_{\alpha \gamma}$ between species $\alpha$ and $\gamma$ through the
first equation of the \bgy hierarchy equation,
\begin{multline}
  \label{BGY}
  \frac{d}{dx} \left( \ln \roax \right) = - \beta \frac{d}{dx} \left
  ( e_\alpha \Phi (x) + \ead \Vself (x) \right) \\
 - \beta e_\alpha \int
  d\vr' \, \left( \sum_\gamma e_\gamma \rho_\gamma (x') \hagrrp
  \right) \frac{\partial \vw}{\partial x} ( \vr';\vr) 
\end{multline}
In \eqref{BGY}  $\Vself(x)$ is the self-energy \eqref{defVself} due to the dielectric
response of the wall, while $\Phi(x)$ is the electrostatic potential
created by the charge density profile $\sum_\gamma e_\gamma \rho_\gamma (x)$. 
$\Phi(x)$ obeys Poisson equation
\begin{equation}
  \label{PoissonPhi}
  \Delta \Phi (x) = - 4 \pi \sum_\gamma e_\gamma \rho_\gamma (x)
\end{equation}
$\Phi$ is uniform in the bulk, since  a fundamental property of
Coulomb systems is the local neutrality relation obeyed by the bulk densities
\begin{equation}
  \label{localneutr}
  \soma \ea \roab = 0
\end{equation}
for any value of the Coulomb coupling parameter. Thus, if  we redefine 
$\Phi(x)$ as the difference between the electrostatic potential created by the
charge density and its bulk value, $\Phi (x)$ tends 
 to zero when $x$ goes to $+\infty $.

We recall that, in the case $\ew =1$, where there is no image
forces, the density is merely uniform in the zero-coupling limit. 
In a plasma with
no charge symmetry the potential drop $\Phi (x)$, which is determined
from the charge density profile $\sum_\gamma e_\gamma \rho_\gamma (x)$ through Poisson
equation \eqref{PoissonPhi}, does not vanish. Moreover, in the weak coupling regime, it is
of the same order as the pair-correlation contribution to $d \roa / dx $
in the \bgy equation, as shown {\it a posteriori} by our explicit calculations
displayed in Section 3. By a mean-field scheme Guernsey \cite{Guer70} closed 
the second \bgy equation in a weak-coupling
limit and found that the  zeroth-order pair
correlation is calculated with uniform densities in a semi-infinite space. Thus
he obtained  coupled equations for $d \roa / dx$ and $\Phi (x)$ and calculated
$\soma \ea \roa (x)$ as a double integral.
(The resolution was not performed for every $\roa(x)$ in the case 
$\Phi (x) \neq 0$  though it might have been done.) The density profiles 
$\roa (x) $ were studied only in the case of a symmetric 
two-component plasma.
The
first-order correction to their bulk value in the weak-coupling regime
was calculated by Jancovici \cite{Janco82I} as follows. Because of the charge symmetry specific to
this system, the charge density profile and subsequently the
electrostatic potential difference with the bulk $\Phi (x)$ vanish at
any distance from the wall. Then the gradient of the density $\roa (x) $
of species  with charge $\ea$ given by the first \bgy equation \eqref{BGY}
 is determined at leading order only by the zeroth-order pair
correlation ; the latter  is calculated in a mean-field approximation for the direct
correlation function with the same
result as that found by Guernsey \cite{Guer70}. 

When $\ew\neq 1$ the methods introduced in the case $\ew =1$ cannot be
generalized straighforwardly, because the fast variation of the density in the vicinity of the
wall prevents one from using mere linearizations. In order to circumvert this
difficulty we introduce the following scheme.

\subsection{Mean-field Ursell function}

In the first \bgy equation  \eqref{BGY} $\sum_\gamma \eg \rog (x') \hagrrp $ 
is the excess charge density of the screening cloud around a charge $\ea$ located 
at $\vr$, the excess charge being calculated with respect to the charge density profile
$\sum_\gamma \eg \rog (x)$. The electrostatic potential created at
$\vr"$ by the charge $\ea$ and its screening cloud is 
\begin{equation}
  \label{expVexc}
  \Phi_{{\rm exc}, \alpha} (\vr; \vr'') = \ea \vw (\vr; \vr'') + \int d\vr'
  \, \left( \sum_\gamma \eg \rog (x') \hagrrp
  \right) \vw (\vr'; \vr'')
\end{equation}
A mean-field approximation amounts to assuming that 
\begin{subequations}
\label{MF}
\begin{align}
    \Phi_{{\rm exc}, \alpha}^{{\scriptscriptstyle MF}} (\vr; \vr'') & 
 = \ea \fcm  (\vr; \vr'') \\
    h_{\alpha \gamma}^{{\scriptscriptstyle MF}} (\vr; \vr') &
 = - \beta \ea \eg \fcm (\vr; \vr')
\end{align}
\end{subequations}
(At leading order in the parameter $\epsd$, the long-range Coulomb interaction
prevails and  the short-distance repulsion
between particles is not involved in \eqref{MF}).
By inserting these approximations into the definition \eqref{expVexc},
we obtain the well-known mean-field equation
\begin{equation}
  \label{eqphiCM}
  \fcm (\vr; \vr'') = \vw (\vr; \vr'') - \beta \int d\vr' \, \left
  ( \sum_\gamma \eg^2 \rog (x') \right) \vw(\vr; \vr') \fcm (\vr'; \vr'')
\end{equation}

Then the mean-field approximation of the
integral in \eqref{BGY} proves to be equal to
\begin{equation}
  \label{coucouu}
  - \frac{\beta \ead}{2} \frac{\partial}{\partial x} \left[ \fcm - \vw
    \right] (\vr; \vr)
\end{equation}
Indeed, the integral 
in \eqref{BGY} can be rewritten by means of the trick involving the Dirac distribution
$\delta (\vr-\vr'')$,
\begin{equation}
  \label{}
  \int d\vr' \, g(\vr; \vr') \frac{\partial f}{\partial x} (\vr'; \vr) =
  \int d\vr'' \, \delta(\vr - \vr'') \frac{\partial}{\partial x} \left
  ( \int d\vr' \, g(\vr''; \vr') f(\vr'; \vr) \right)
\end{equation}
Moreover, according to \eqref{pois} and \eqref{eqphiCM}, $\fcm (\vr; \vr')$ is
the Green function of the operator 
$\Delta_{\vr'} -4\pi\beta \sum_\gamma \eg^2 \rog (x')$. Since the latter operator
is self-adjoint, the real function $\fcm (\vr; \vr')$, as well as $\vw (\vr;
\vr')$,  is symmetric under
exchange of $\vr$ and $\vr'$ when $\vr$ and $\vr'$ are in the same region. Then
\eqref{eqphiCM} can be used. Since a symmetric function
$h(\vr;\vr')=h(\vr';\vr)$ obeys the identity
$\partial[h(\vr;\vr')]/\partial x\vert_{\vr=\vr'}
=(1/2)\partial[h(\vr;\vr)]/\partial x$, we get the result \eqref{coucouu}.

Finally, according to the definition \eqref{defVself} of $\Vself (x)$ and since
$\roax$ tends to $\roab$ when $x$ goes to 
$+\infty$, the mean-field density profile $\roacm (x)$ proves to read
\begin{equation}
  \label{rocmm}
  \roacm (x) = \theta (x-b) \roab \exp \left[ - \beta \ead \Vselfcm (x) - \beta \ea
  \Pcm (x)  \right]
\end{equation}
with
\begin{equation}
  \label{fcmmvb}
  \Vselfcm (x) \equiv \frac{1}{2} \left( \fcm - \vb \right) (\vr,
  \vr) - \lim_{x \rightarrow + \infty} \frac{1}{2} \left( \fcm - \vb
  \right) (\vr, \vr) 
\end{equation}
Meanwhile the coupled equation for the electrostatic potential $\Pcm(x)$ is
\eqref{PoissonPhi} where $\rho_{\gamma}(x)$ is replaced by 
$\rho_{\gamma}^{{\scriptscriptstyle MF}}(x)$.
In \eqref{rocmm} the argument in the exponential may be interpreted as $\beta$ times the
work given by an operator to the system in order to put a charge $\ea$
into the Coulomb fluid at $\vr$, make it cross the potential drop
$\Pcm (x)$ from $x$ to $+\infty$ and then get it back from the
bulk. In the following, we set
\begin{equation}
  \label{defVscself}
  \zacm (x) \equiv \theta (x-b) \roab \exp \left[ - \beta \ead \Vselfsc (x) \right]
\end{equation}

%%%%%%%%%%%%%%%%%%%%%%%%%%%%%%%%%%%%%%%%%%%

\subsection{Linearization of the $\Phi(x)$ contribution}

Explicit calculations can be performed if the contribution from 
$\Pcm (x)$ to
\begin{equation}
  \label{}
  \roacm (x) = \zacm (x) e^{- \beta \ea \Pcm (x)}
\end{equation}
is linearized, 
\begin{equation}
  \label{rolinu}
  \rocmlin (x) = \zacm (x) \left[ 1 - \beta \ea \Pcmlin (x) \right]
\end{equation}
Such a linearization is allowed only if $\Pcm (x)$ does not become infinite in the
vicinity of the wall. This is indeed the case,  because $\Pcm (x)$
is created by the charge density and the latter has no singularity thanks to the
hard-core repulsion from the wall. The absence of divergency in $\Phi(x)$ near
the wall  is also checked  in the systematic approach of 
Paper II. In the following we will show that the screened self-energy 
$\Vselfcm(x)$
diverges when $x$ goes to zero and no linearization can be performed for it.

By inserting the linearized mean-field expression for the densities into  Poisson
equation \eqref{PoissonPhi}  we find that  
\begin{equation}
  \label{sixtreize}
  \left[ \Delta_{\vr} - 4 \pi \beta \sum_\gamma \egd \zgcm (x) \right]
  \Pcmlin (x) = - 4 \pi \sum_\gamma \eg \zgcm (x)
\end{equation}
\eqref{sixtreize} can be viewed as some kind of partially linearized Poisson-Boltzmann
equation in an inhomogeneous case. As a consequence of \eqref{sixtreize}, 
\begin{equation}
  \label{}
  \Pcmlin (x) = \int d\vr' \, \fcmlin (\vr', \vr) \sum_\gamma \eg
  \zgcm (x') 
\end{equation}
where  $\fcmlin (\vr, \vr')$ is a Green function solution of 
\begin{equation}
  \label{sseize}
  \left[ \Delta_\vr - \kd^2 \left( 1 + U(\vr) \right) \right] \fcmlin
  (\vr, \vr') = - 4 \pi \delta \left( \vr - \vr' \right)
\end{equation}
with 
\begin{equation}
  \label{sixseize}
  U(\vr) = \frac{4 \pi \beta}{\kd^2} \sum_\gamma \egd \left[ \zgcm
  (x) - \rogb  \right]
\end{equation}
$\fcmlin$ is the solution of \eqref{sseize} which satisfies the same boundary conditions 
as $\vw (\vr, \vrp)$.

%%%%%%%%%%%%%%%%%%%%%%%%%%%%%%%%%%%%

\subsection{Solution of the inhomogeneous Debye-H\"uckel equation at leading order}

A formal solution of the inhomogeneous Debye-H\"uckel equation \eqref{sseize}
is given in Paper II. An $\epsd$-expansion is devised and we show that
\begin{equation}
  \label{sdh}
  \fcmlin (\vr,\vr')= \kd \ftz (\kd \vr, \kd \vr') \left[ 1 + 
  {\mathcal O}_{{\rm exp}} (\epsd) \right] 
\end{equation}
  In \eqref{sdh} $\Oexpeps$ denotes a function of order $\epsd$ -- possibly
  multiplied by some power of $\ln \epsd$ -- which 
decays exponentially fast at large distances over a scale $\kd^{-1}$ and which remains bounded
 by a function of $\beta e^2 / b$ for all $x$ larger than the closest approach distance $b$ to the wall. 
$\kd \ftz$ is the solution of the homogeneous Debye-H\"uckel equation
\begin{equation}
  \label{DHhom}
  \left[ \Delta_{\vr} - \kd^2 \right] \kd \ftz \left(\kd \vr, \kd \vrp \right) =
- 4 \pi \delta (\vr - \vrp)
\end{equation}
with the same electrostatic boundary conditions as $\vw(\vr, \vrp)$ : $\ftz$ is continuous in
all space and
\begin{equation} 
\lim_{x \rightarrow 0^-} \ew \frac{\partial \ftz}{\partial x}
(\kd \vr, \kd \vrp) = \lim_{x \rightarrow 0^+} \frac{\partial \ftz}{\partial
  x} (\kd \vr, \kd \vrp)
\label{boundcond}
\end{equation}
Equation \eqref{sdh} also 
holds for $\fcm$ defined in \eqref{eqphiCM}
and which obeys equation \eqref{sseize} where $ \zgcm (x)$ in $U(\vr)$
is replaced by $\rog (x)$. Thus the screened self-energy defined in 
\eqref{fcmmvb} is also determined at leading order in $\epsd$.

Eventually, the density profile at leading order in $\epsd$ is given by 
\eqref{defVscself} 
and \eqref{rolinu} with
\begin{equation}
  \label{toutou}
  \Vselfsc (x)= \frac{1}{2} \kd \left[\ftz - \ftb \right] (\kd \vr, \kd \vr)
\end{equation}
and
\begin{equation}
  \label{phiphi}
\Pcmlin (x)= 
\int d \vrp \kd 
\ftz (\kd \vr, \kd \vrp) \somg 
\eg 
\zgcm (x') 
\end{equation}
In \eqref{toutou} $\ftb$ denotes the solution of the homogeneous 
Debye-H\"uckel equation in the bulk, namely the solution 
which vanishes when $|\vr-\vrp|$ becomes infinite,  as well as  $\vb (\vr, \vrp)$
defined in \eqref{defvb}.

%%%%%%%%%%%%%%%%%%%%%%%%%%%%%%%%%%%%%%%%%%%%

\section{Explicit density profiles}
\label{section3}

\subsection{Solution of the homogeneous Debye equation}

 Equation \eqref{DHhom} can be solved because it is changed into a second-order
 differential equation by  taking the 
Fourier transform in the direction parallel to the wall, 
\begin{equation}
  \label{defFour}
  \ftz (\xt, \xt', \vq) = \int d\vyt \, e^{i \vq. \vyt} \ftz (\xt, \xt', \vyt) 
\end{equation}
In \eqref{defFour} we have used the dimensionless variables $\xt = \kd x$ and
$\vyt = \kd \vy$, where $\vy$ is the projection of $\vr$ onto a plane parallel to the
 wall. As recalled in Paper II, since $\kd \ftz$ obeys the same boundary conditions as $\vw$, for $x>b$
 and $x'>b$
\begin{multline}
 \label{defphi10}
  \ftz(\xt,\xt ',\vq;\bt,\dw) =  \ftb (|\xt - \xt '|, \vq) \\
          + \Zv \hthw (\xt + \xt '- 2\bt ; \vq)
\end{multline}
with 
\begin{equation}
\ftb (|\xt- \xt '|,\vq) = \frac{2 \pi}{\rac}
\, e^{- |\xt - \xt '| \rac}   
\label{valuephib0}  
\end{equation}
\begin{equation}
  \label{valuehop}
 \hthw (\xt + \xt' - 2\bt ;\vq) =  
 \frac{2 \pi}{\rac} \, \frac{\rac -
  |\vq|}{\rac + |\vq|} e^{- (\xt + \xt '-2 \bt) \rac}
\end{equation}
and
\begin{equation}
  \label{defZ}
  \Zv \equiv \frac{ 1 - \dw e^{-2 q \bt } \left[ \rac + \vert\vq\vert
  \right]^2}{1 - \dw e^{-2 q \bt } \left[ \rac - \vert\vq\vert
  \right]^2} 
\end{equation}

%%%%%%%%%%%%%%%%%%%%%%%%%%%%%%%%%%%%%%%%%%

\subsection{Screened self-energy}

The screened self-energy \eqref{toutou} can be written as 
\begin{equation}
  \label{relVL}
  \beta \ead \Vselfcm(x)= \epsa L (\xt - \bt; \bt, \dw)
\end{equation}
with $\epsa \equiv (1/2) \beta \ead \kd$ and
\begin{equation}
  \label{}
  L(\xt - \bt;\bt, \dw) \equiv 
  \int \frac{d^2 \vq}{(2 \pi)^2} 
 \left[ \phit^{(0)}
   (\xt,\xt,\vq; \bt ,\dw) - \phit_{{\rm \scriptscriptstyle B}} (\xt,\xt,\vq) \right]
\end{equation}
According to \eqref{defphi10}
\begin{equation}
  \label{defLt}
    L(u;\bt, \dw) =  \int \,
   \frac{d^2 \vq}{(2 \pi)^2}  \, \Zv  
   \times \htpzv(2 u ; \vq)
\end{equation}
where $\htpzv$ is given in \eqref{valuehop}.
By using the change of variable $t = \rac$  we get 
\begin{equation}
  \label{valueL}
  L(\xt - \bt; \bt, \dw ) = \int_1^{+\infty} dt \, \frac{ 1 - \dw \left[ t
  + \racm \right]^2 e^{-2 \bt \racm}}{\left[ t + \racm \right]^2 - \dw
  e^{-2 \bt \racm}} e^{ - 2 (\xt - \bt) t}
\end{equation}

The successive changes of variables $t = t'+1$ then $t'= v/
 \left[ 2(\xt-\bt) \right] $
allow to show that 
\begin{equation}
  \label{ch}
  L(\xt - \bt; \bt, \dw) \underset{\xt \rightarrow + \infty}{\sim}
  \frac{e^{-2(\xt - \bt)}}{ 2(\xt - \bt)} 
\end{equation}
If $\bt \neq 0$ the integrand in \eqref{valueL} behaves as $1/t^2$ times
$\exp \left[ - 2 (\xt - \bt ) t \right] $ when $t$ goes to $+ \infty$
and $L(\xt-\bt, \bt, \dw)$ is finite for all values of $\xt$ even when
$\xt$ approaches $\bt$. If $\bt = 0 $ the integrand vanishes as $\exp(-
2 \xt t)/t^2$ for large $t$ when $\dw =0 $ but it behaves as $-\dw
\exp(-2 \xt t)$ if $\dw \neq 0$. Subsequently for $\bt = 0$, the
integral diverges at $\xt = \bt =0$ when $\dw \neq 0$. By subtracting
the dangerous asymptotic behaviour $I_{{\rm as}}(t) = - \dw \exp \left[ - 2
  \xt t \right] $ from the integrand of $L$ and by performing
$\int_1^{+\infty} dt \, I_{{\rm as}} (t)$ we get
\begin{equation}
  \label{decompL}
  L(\xt - \bt; \bt, \dw) = - \dw \frac{e^{-2 \xt}}{2 \xt} + \Lb (\xt ; \bt
  , \dw)
\end{equation}
where $\Lb (\xt ; \bt, \dw)$ remains finite even when $\xt = \bt = 0$.

%%%%%%%%%%%%%%%%%%%%%%%%%%%%%%%%%%%%%%%%%%

\subsection{Electrostatic potential drop}

In order to calculate the electrostatic potential drop \eqref{phiphi}, we notice
that 
\begin{equation}
\label{add}
\widetilde{\phi}^{(0)} (\xt, \xt', \vq = {\bf 0};\bt)  
= 2\pi\,[e^{-  | \xt - \xt' |} + e^{- (\xt+ \xt'-2\bt)}]
\end{equation}
and we rewrite $\zacm (x)$ as
\begin{equation}
  \label{}
  \zacm(x) = \roab \left[1 + w_0 (\xt; \epsa, \dw) \right] 
  \left[1 - \epsa \Lb (\xt; \bt, \dw) \right]
\end{equation}
with 
\begin{equation}
  \label{}
  \wza \equiv \exp \left[ \dw \frac{\beta e_{\alpha}^2}{4 x } e^{-2 \kd x} \right] - 1
\end{equation}
According to the bulk neutrality relation \eqref{localneutr}, 
the constant
term in $\zacm (x)$ gives a vanishing contribution to \eqref{phiphi}. 
$w_0 (\xt; \epsa,\dw)$ proves 
to contribute from order $\epsd\ln\epsd$ to the integral in \eqref{phiphi}.
Indeed, let us  consider a function 
$f(\xt';\xt,\bt)$ which is bounded for all $\xt'\geq 0$ and $\bt\geq 0$.
 If $b>\kd^{-1}$, then for all $x>b$ $\beta e_{\alpha}^2/x<\kd \beta
e_{\alpha}^2$
and 
\begin{align}
\label{trotrodeux}
 \kd\int_{b}^{+\infty} d x' \, & w_0(\xt'; \epsa,\dw) f(\xt';\xt,\bt)\nonumber\\ 
& \mathop{\sim}_{\epsa \rightarrow 0} 
\kd\int_{\bt}^{+\infty} d\xt' \,  
\frac{\dw\beta e_{\alpha}^2}{4 \xt'}e^{-2\xt'}
 f ( \xt ';\xt,\bt) 
\end{align}
In the case $b\ll \kd^{-1}$, 
let us introduce the length $l$ such that  
$\beta e_{\alpha}^2\ll l\ll
\kd^{-1}$. For all $x$ in the range $b<x<l$, $\kd x\ll 1$,
while, for all $x>l$, $\beta e_{\alpha}^2/x\ll 1$.
Then at leading order in $\epsd$
\begin{align}
  \label{trotro}
 \kd\int_{b}^{+\infty} & d x' \, w_0(\xt'; \epsa,\dw) f(\xt';\xt,\bt)\nonumber \\
&\mathop{\sim}_{\epsa \rightarrow 0} \,\,
\lim_{l/\beta e_{\alpha}^2\rightarrow+\infty}\,\,
\lim_{\kd l\rightarrow 0}\nonumber\\
&\qquad\left\{ 
 \kd\int_{b}^{l}
dx' \, \left[ \exp\left(\frac{\dw\beta e_{\alpha}^2}{4x'}\right)-1\right]
f(\xt'=0;\xt,\bt=0)\right. \nonumber\\ 
&\qquad\left. + \kd\int_{\lt}^{+\infty} d\xt' \,  
\frac{\dw\beta e_{\alpha}^2}{4 \xt'}e^{-2\xt'}
 f ( \xt ';\xt,\bt=0) 
\right\}\nonumber\\
&\times \left[1+\cO(\bt)\right]
\end{align}
where $\cO(\bt)$ denotes a term of order $\bt$.
After the change of variable $x'=x'_1\beta e_{\alpha}^2$, the first integral in
\eqref{trotro} proves to be of order $\epsa$, as well as the second integral.
Both integrals have a logarithmic dependence upon $l$ and the respective 
$\ln (l/\beta\ead)$ and $\ln(\kd l)$ terms combine so that the sum of the two
integrals starts at order $\epsd\ln\epsd$.

Eventually,  we get for $\Pcmlin (x)$,
denoted by $\Phi(x)$ in the following, 

\begin{equation}
  \label{vtotcmlin}
  \Phi (x) = - \frac{2 \pi \beta}{\kd} \somg \rog \eg^3 M_\gamma 
 (\kd x;  \epsg, \kd b, \dw)\times
  \left[1+{\mathcal O}_{{\rm exp}} (\epsd) \right]
\end{equation}
where  ${\mathcal O}_{{\rm exp}} (\epsd)$ is defined after 
\eqref{sdh}. In \eqref{vtotcmlin}
$M_\gamma =\bar{M} +[M_\gamma -\bar{M}]$ with 
\begin{equation}
  \label{exprMb}
  \bar{M} (\xt; \bt, \dw) = \frac{1}{2} \int_{\bt}^{+\infty} \, du' \,
 \left[ e^{-|\xt-u'|} + e^{-(\xt+u'-2\bt)} \right] \Lb (u';\bt,\dw) 
\end{equation}
and $M_\gamma - \bar{M}$ is the $\epsd$-expansion at orders $\ln
\eps_\gamma$ and $(\eps_\gamma)^0$ of the integral 
\begin{equation}
  \label{ctu}
  - \frac{1}{2} \frac{1}{\eps_\gamma} \int_{\bt}^{+\infty} \, du' \,
  \left[ e^{-|\xt -u'|} + e^{-(\xt +u'-2\bt)} \right] w_0(u';
  \eps_\gamma, \dw)
\end{equation}

%%%%%%%%%%%%%%%%%%%%%%%%%%%%%%%%%%%%%%%%%%%%%%%%%%%%%%%%%%%%%%%%%%%

\subsection{Density profile}

The previous expressions are inserted in \eqref{rolinu} 
with the result 
\begin{align}
  \label{rhoexp}
  \roa(x) = & \roab \theta(x-b)   \exp \left( \dw \frac{\beta \ead}{4x} 
  e^{-2 \kd x} \right) \nonumber\\
 & \times \left\{ 1 \rule{0mm}{8mm} 
 - \frac{1}{2} \beta \kd \left[ \rule{0mm}{7mm}
  \ead \Lb 
 \left( \kd x; \kd b, \dw \right) \right.  \right. \nonumber \\
& \phantom{\times \left\{ 1 - \frac{1}{2} \beta \kd \left
  [ \rule{0mm}{7mm} \right. \right. } 
\left. \left. - e_\alpha \frac{4
  \pi \beta }{\kd^2} \sum_\gamma \rogb e_\gamma^3 M_\gamma 
(\kd x;\eps_\gamma, \kd b, \dw) \right]  
  \rule{0mm}{8mm} \right\}  
\end{align}
where $\epsg = (1/2) \beta \egd \kd$ . $M_\gamma$ is defined in 
\eqref{exprMb} and \eqref{ctu}, while $\Lb$ is given by \eqref{valueL} and \eqref{decompL}.  $\Lb (\xt; \bt, \dw)$ decreases 
exponentially fast as $\exp(- 2 \xt)/ 2 \xt$ when $\xt$ goes
to $\infty$, while $M_\gamma$ decays only as $\exp(-\xt)/\xt$.

We give more explicit  formulae in the regime
\begin{equation}
  \label{condkb}
  \eta \equiv \kd b \ll 1
\end{equation}
whatever the value of $\beta e^2 / b$ may be. When $\epsd\ll 1$, according to
\eqref{blibli} and \eqref{defkdbis}, the mean interparticle distance  $a$ is smaller than
$\kd^{-1}$, $a< \kd^{-1}$ and the
condition \eqref{condkb} will be fulfilled if $b \ll a$ -- for instance 
if $b$ is of the same magnitude as
the hard-core diameter  of charges which itself is far smaller
than $a$. 
$\Lb (\xt; \kd b, \dw)$ is bounded for every $x$, even when $\eta=\kd b$ vanishes,
and it can be expanded in powers of $\eta$. 
According to \eqref{valueL} and \eqref{decompL}, $\Lb (\xt ; \eta= 0,
\dw)$ is directly given by 
\begin{equation}
  \label{Lbar}
  \Lb(\xt;\eta = 0, \dw) = \int_1^{+\infty} \, dt \, e^{-2 t \xt } \frac
  { (1-\dw^2)}{(t+\sqrt{t^2-1})^{2} - \dw}
\end{equation}
On the other hand
\begin{multline}
  \label{decompM}
  M_\gamma (\xt; \eps_\gamma,\eta=0, \dw) \\
  = \bar{M} (\xt; \eta=0, \dw) + (\ln
  \eps_\gamma) \dw \frac{1 }{2} e^{-\xt} 
 - \dw I_\gamma (\xt;\dw, \beta e_\gamma^2 / b)
\end{multline}
where
\begin{equation}
  \label{expMb}
  \Mb(\xt;\eta=0, \dw) =  \int_1^{+\infty} \, dt \, \left[ \frac{ 
 e^{-2t \xt} -2t e^{-\xt} }{1-(2t)^2} \right] 
 \frac{1-\dw^2}{\left( t +\sqrt{t^2-1}
  \right)^2 - \dw}
\end{equation}
while, according to \eqref{trotro},
\begin{align}
  \label{Igamma}
  I_\gamma (\xt; \dw, \beta e_\gamma^2 / b) = &- \frac{1}{4} e^{-\xt}
  \left\{ 2\left[ 
  A \left( \frac{\beta e_\gamma^2 \dw}{4b} \right) +
  \ln \left( \frac{|\dw|}{2} \right)
  + 2 \C - 1 \rule{0mm}{6mm} \right]
+ \ln 3 \rule{0mm}{7mm} \right\}\nonumber\\
&+ \frac{1}{4}e^{-2 \xt}
  \left[ \rule{0mm}{5mm} e^{\xt} \Ei (-\xt) - e^{3\xt} \Ei (- 3 \xt) \right] 
\end{align}
In \eqref{Igamma} $\C$ is the Euler constant, $\Ei(-u)$ denotes the 
Exponential-Inte\-gral function defined for 
$u>0$ as $\Ei(-u) \equiv - \int_u^{+\infty} dt \, \exp(-t)/t$ and
\begin{equation}
  \label{cddefA}
  A (u) \equiv \frac{1}{u} \left[ e^{u} - 1 \right] - \Ei \left
  ( u \right)
\end{equation}
$A(u)$ arises from the integration of $\exp(-\beta \eg^2 \dw / 4u')-1$ in
\eqref{ctu}.

%%%%%%%%%%%%%%%%%%%%%%%%%%%%%%%%%%%%%%%%%%%%%%%%%%%%%%%%%%%%%%%%%%%%%%%%%%%%%%%

\subsection{Interpretation : competition between three effects}

The profile density is ruled by the competition between three kinds of
effective interactions, as exhibited by rewriting the density profile
 by means of \eqref{rocmm}, \eqref{relVL} 
and \eqref{decompL} 
with the result
\begin{align}
  \label{fondrho}
  \roa (x) = \roab & e^{\dw \beta \ead (e^{- 2 \kd x} / 4x)} \nonumber \\
 & \times \left\{ 1 - \beta \ead \frac{\kd}{2} \Lb (\kd x; \kd b, \dw) -
 \beta \ea \Phi (x) + \Oexp (\epsd^2) \right\} 
\end{align}
where  ${\mathcal O}_{{\rm exp}} (\epsd^2)$ is defined after 
\eqref{sdh}.
The interpretation of \eqref{fondrho} is the following. 

First, $\ea \Phi (x)$ is the interaction between a charge $\ea$ and the charge
profile density in the electric layer. The other two interactions, which are
proportional to $\ead$, are the two parts of the screened self-energy.

Second, $\exp (\dw \beta \ead \exp(- 2 \kd x)/4x)$ is the effective Boltzmann factor
associated with the part of the screened self-energy created by the
electrostatic response of the wall. The effect of the corresponding
attractive ($\dw >0$) or repulsive ($\dw < 0$) interaction with the
wall cannot be linearized. Indeed, when the dielectric wall is
repulsive, the density vanishes as $\exp ( - |\dw| \beta \ead /4b)$ when
$b$ goes to zero. Since the hard-core repulsion is spurious when 
$\dw < 0$, we can set $b=0$ and 
\begin{equation}
  \label{}
\roa (x=0, \dw < 0) = 0    
\end{equation} 
On the contrary when $\dw >0$, the density blows up as 
\begin{equation}
  \label{}
  \roa (x=b, \dw) \underset{b \rightarrow 0}{\sim} \roab \exp 
 \left\{ \frac{\dw
  \beta \ead e^{-\kd b}}{4b} \right\} \left[ 1 + \cO (\epsd) \right]
\end{equation}

Third, the
part of the screened self-energy which exists even in the absence of any
electrostatic property of the wall (namely even when $\dw = 0$) is
$(\kd/2) \Lb(\kd x; \kd b, \dw)$; it arises from the ``geometric'' repulsion
caused by the mere presence of  the wall. Indeed, by deforming the screening
cloud (with a net charge $- \ea$) surrounding any charge $\ea$, the wall 
hinders the stabilizing
effect of Coulomb interactions, as exhibited clearly in 
subsection 5.3
about the plain hard wall.

%%%%%%%%%%%%%%%%%%%%%%%%%%%%%%%%%%%%%%%%%%%%%%%%%%%%%%%%%%

\subsection{Limiting case of the OCP}

The density profile for a one-component plasma (OCP) can be derived from
the expression obtained for a two-component plasma by the following
trick already tested in Refs. \cite{Alas&Pere96,Cornu98II}. In 
order to describe a OCP where moving particles carry a positive charge
$e$ in a neutralizing uniform background with density $\rho$, we
start from a two-component plasma with $e_+ = e$ and $\rho_+ = \rho$ and
we take the limit where $e_- / e_+$ vanishes while $e_- \rho_- = - e_+
\rho_+ $ is kept fixed. 

\raggedbottom

The general expression \eqref{rhoexp} tends to the limit
\begin{multline}
  \label{rhoocp}
  \rocp= \rho \, \theta (x-b) \exp \left[ \frac{ \dw \beta e^2 e^{- 2 \kd
  x}}{4x} \right] \\
 \times \left\{ 1 - \frac{1}{2} \beta \kd e^2 \left[ \Lb (\kd
  x; \kd b, \dw) - \Mocp (\kd x; \beta \kd e^2/2, \kd b,  \dw) \right]
  \right\} 
\end{multline}
where $\kd = \sqrt{ 4 \pi \beta \rho e^2}$, $\Lb$ is defined by \eqref{valueL}
and \eqref{decompL}, while $\Mocp$ is equal to the function $M_\gamma$ defined
in \eqref{exprMb} and \eqref{ctu} with $\epsg$ replaced by $\kd \beta e^2 /2$. 
According to \eqref{vtotcmlin}, the potential difference with the bulk is
\begin{equation}
  \label{}
  \Phi_{\ssocp}(x) = 
 - \frac{1}{2} \kd e \Mocp (\kd x; \beta \kd e^2 /2, \kd b, \dw)
\end{equation}

We recall that \eqref{rhoocp} can also be rewritten as
\begin{equation}
  \label{rhoocpbis}
  \rocp (x) = \rho \, \theta (x-b) \, e^{- \beta e^2 \Vselfsc (x)} \left[ 1 -
  \beta e \Phi_{\ssocp} (x)  \right]
\end{equation}
where $\Vselfsc (x)$ is given in \eqref{undix}. Moreover, according to
\eqref{add}, \eqref{phiphi}  may be rewritten in the case of the OCP as 
\begin{multline}
  \label{generalphi}
  \frac{1}{2} \int_{\bt}^{+\infty} d \xt' \left[ e^{- \beta e^2 \Vselfsc
  (\xt')} - 1 \right] \left[ e^{- |\xt - \xt'|} + e^{- (\xt + \xt' - 2
  \bt)} \right] \\
  = \beta e \Phi_{\ssocp} (x; \bt) 
 + {\mathcal O}_{{\rm exp}} (\epsd^2) 
\end{multline}

%%%%%%%%%%%%%%%%%%%%%%%%%%%%%%%%%%%%%%%%%%%%%%%%%%%%%%

\subsection{Comparison with previous results}

The expression \eqref{rhoocp} for the OCP 
is valid at any distance from the wall and
for any value of $\dw$ and $\kd b$. Let us see how it can be compared
with the expression (3.21) in Ref.\cite{Alas83} obtained in the case
$\dw < 0$ and $b=0$, and for distances $ x \gg \beta e^2$. (The
expressions in Ref.\cite{Alas83} will be denoted  by a superscript *.) 

Alastuey starts from the BGY equation \eqref{BGY} and directly replaces
the Ursell function by its Debye approximation 
$- \beta e^2 \kd \ftz (\kd \vr, \kd \vrp; \dw)$. The equation (3.17) in Ref.\cite{Alas83} -- which
is analogous to our equation \eqref{sixtreize} -- involves the screened
self-energy calculated at leading order in $\epsd$ -- as in the present paper --
and
given by
%in \eqref{idL}
\begin{equation}
  \label{selfap}
  \Vselfsce (x) = \epsd L (\kd x;  \dw)
\end{equation}

However the equation (3.17) in Ref.\cite{Alas83} is solved only for
distances $x \gg \beta e^2$ which are large enough to allow one to
replace $U(\vr)$, defined in our equation \eqref{sixseize}, by zero. 
The corresponding approximated $\Phi_{\ssocp}^*$ is solution of 
\begin{equation}
  \label{eqapproch}
  \left[\frac{d^2}{dx^2} - \kd^2\right] \Phi_{\ssocp}^* (x)= - 
 \frac{\kd^2}{\beta e} e^{ - \beta e^2 \Vselfsce (x)}
 \qquad x \gg \beta e^2
\end{equation}
while, according to the Poisson equation which
relates $d^2\Phi_{\ssocp}^*/dx^2$ and the charge density 
$e \left[ \rho(x) -   \rho \right]$,
\begin{equation}
  \label{rhoocpap}
  \rho_{\ssocp}^* (x)= \rho \left[ e^{ - \beta e^2 \Vselfsce (x) } - \beta
  e \Phi_{\ssocp}^* (x) \right] \qquad x \gg \beta e^2
\end{equation}
(\eqref{rhoocpap}is the formula (3.21) given in Ref. \cite{Alas83}.)
According to the result \eqref{sdh} of our analysis of the non
approximated  equation \eqref{sseize} (see Paper II), the expression of
$\Phi_{\ssocp}^*$, solution of \eqref{eqapproch} and  given in Eq.(3.20) of 
Ref.\cite{Alas83}, coincides with
our formula \eqref{generalphi} valid at any distance. 
On the other hand, the result \eqref{rhoocpap} does coincide with the 
$x \gg \beta e^2$ limit of the expression \eqref{rhoocpbis} which is valid
for any $x$. 

Our ability to handle with all distances relies on two
progresses with respect to the approach of Ref.\cite{Alas83}. First, we
extract from the expression of $\ead \Vselfsce (x)$ the part which
diverges as the bare self-energy $\dw \ead / (4x)$ when $x$ approaches
the wall and which enforces the vanishing of the density at the wall when
$b=0$. Second, we are able to disentangle this short-range effect from
the long-range exponential screening through our systematic method of
expansion introduced in Paper II for the solutions of the inhomogeneous
Debye-H\"uckel equations at stake.

%%%%%%%%%%%%%%%%%%%%%%%%%%%%%%%%%%%%%%%%%%%%%%%%%%%%%%%%%%%%%%%%%%%%%%%%%

\section{Generic global properties}
\label{section4}

\subsection{Potential drop}

First we recall that in the generic case the local neutrality valid in the bulk is destroyed
near the wall, 
\begin{equation}
  \label{cnez}
  \sum_\alpha \ea \roax \neq 0
\end{equation}
and there appears an electric layer which is responsible for a potential
drop $\Phi (x)$ between each point and the bulk. The expression of $\Phi(x)$ is
given in \eqref{vtotcmlin}.
$\Mb$ in the decomposition of $M_\gamma$
 is a positive function, whereas $M_\gamma - \Mb$ may have any sign. Thus
the profile of $\Phi$ depends on the composition $\{ e_\gamma, \rog
\}_{\gamma = 1, \ldots, n_s}$ of the Coulomb fluid and on the value of
$\dw$. 
For instance, when $\kd b \ll 1$,
\begin{multline}
  \label{expdf}
  \Phi (x=0) = - \frac{2 \pi \beta}{\kd} 
 \sum_\gamma \rogb e_\gamma^3  \left\{ 
 \rule{0mm}{8mm} (1-\dw^2) J(\dw) \right. \\
 + \frac{\dw}{2} \left[ \ln \left( 3 \kd \beta \egd 
 |\dw|/4 \right) - 1 + 2\C + A \left( \frac{\dw \beta \egd}{4b}
 \right) \right]\\
 \left. \rule{0mm}{8mm} + \Oc (\epsd;\kd b) \right\}
\end{multline}
In \eqref{expdf} $\cO (\epsd; \kd b)$ denotes a term of order either
$\epsd$ or $\kd b$, and
\begin{multline}
  \label{}
  J(\dw) = \frac{1}{4(1+\dw+\dw^2)} \left\{ -\frac{\pi}{\sqrt{3}} +
  (1+2\dw) \ln 3 \right. \\ 
\left. + \frac{1-\dw^2}{\dw} \ln (1-\dw) +
  \frac{1-\dw}{\sqrt{\dw}} \ln \left( \frac{1+\sqrt{\dw}}{1-\sqrt{\dw}}
  \right) \right\} 
\end{multline}
An example for the profile $\Phi (x) $ is drawn in Fig. \ref{f6}.

 However the local neutrality $\soma \ea \roax = 0$  holds in 
the specific case of a charge-symmetric plasma in a symmetric state, for any 
strength of the Coulomb coupling in the fluid. We
use the following definitions. A charge-symmetric fluid contains 
equal numbers of positively and negatively
charged species and the set of charges is invariant under
inversion of charges. 
In the special case of a two-component plasma made of charges $+e$ and
$-e$, the latter charge symmetry combined with the neutrality relation
\eqref{localneutr} implies that $\rho_+^{\ssB} = \rho_-^{\ssB}$. On the
contrary,  for a
charge-symmetric plasma with at least four species $\roa^{\ssB} \neq
\rho_{- \alpha}^{\ssB}$ in the generic case. However, in some situations
(for instance when two different salts made with monovalent ions are
dissolved in water) the system is prepared in a symmetric state; the
bulk density parameters are chosen to satisfy
\begin{equation}
  \label{defsym}
  \rho_{\alpha}^{\ssB \, {\rm sym}} = \rho_{- \alpha}^{\ssB \, {\rm sym}}
\end{equation}
In such a symmetric state, the symmetry of the Hamiltonian under inversion of
charge signs enforces that at any point $x$
\begin{equation}
  \label{nocharge}
  \sum_\alpha \ea \roa^{{\rm sym}} (x) = 0 
\end{equation}
and, subsequently
%, according to \eqref{sd},
\begin{equation}
  \label{noV}
  \Phi^{{\rm sym}} (x) = 0 
\end{equation}
In a symmetric state, a charge-symmetric Coulomb fluid does not build
any charge density profile or any electrostatic potential difference
with the bulk.

%%%%%%%%%%%%%%%%%%%%%%%%%%%%%%%%%%%%%%%%%%%%%%%%%%%%%%%%%%%%%%

\subsection{Global charge}

A dielectric wall remains globally neutral in
the presence of a Coulombic fluid and may only acquire  macroscopic
multipoles depending on the geometry of the dielectric sample. As a
consequence, we expect \cite{Mart88} that, since $\ew <+\infty$,
\begin{equation}
  \label{sumrule2}
 \int_0^{+\infty} \, dx \, \sum_\alpha e_\alpha \roa (x) = 0 
\end{equation}
 We recall that 
$\Phi (x=0) \neq 0$ means that the dielectric layer carries a nonvanishing
dipole though its net charge is zero.

The global charge of the system calculated at leading order $\epsd$ indeed 
vanishes 
in agreement with \eqref{sumrule2}. The result is readily obtained by writing 
for $\rho_\alpha(x)$ the structure \eqref{rolinu} combined with \eqref{phiphi}
and by using the following property.
According to 
\eqref{trotrodeux} and \eqref{trotro}, if $f(x)$ is an integrable function 
 which is bounded for every $x\geq 0$,
 
 \begin{equation}
  \label{fundprop}
\int_0^{+\infty} dx \, \zacm (x)f(x)
=\roab \int_0^{+\infty} dx \,f(x)
\times \left[1+\cO \left( \epsd \ln \epsd \right)\right]
\end{equation}
 Eqs.  \eqref{rolinu} and \eqref{fundprop} lead to 
\begin{align}
\label{chargelead}
\int_0^{+\infty} dx \,\sum_\alpha e_\alpha \roa (x)
&=\int_0^{+\infty} dx \,\sum_\alpha e_\alpha \zacm (x)\nonumber\\
&-\beta \left(\sum_\alpha e_\alpha^2\roab\right)
\left( \int_0^{+\infty} dx \,\Phi(x)\right) 
\times\left[1+{\mathcal O}(\epsd \ln \epsd)\right]
\end{align}
$\Phi(x)$ is given at leading order $\epsd$ by \eqref{phiphi}, and the property 
\begin{equation}
\label{}
\int_0^{+\infty} dx\int d\vy \,
\kd \ftz(\kd x, \kd x', \kd \vy)={4\pi\over \kd^2}
\end{equation}
implies that
\begin{equation}
  \label{intphi}
 \int_0^{+\infty}  dx \, \Phi(x)
 ={4\pi\over \kd^2}\int_0^{+\infty}  dx'
  \somg \eg \zgcm (x') 
\end{equation}
Combination of \eqref{chargelead} and \eqref{intphi} implies that  
\eqref{sumrule2} is indeed satisfied at leading order in $\epsd$.

%%%%%%%%%%%%%%%%%%%%%%%%%%%%%%%%%%%%%%%%%%%%%%%%%%%%

\subsection{Contact theorem}

Finally, we turn to the so-called contact theorem which gives the
difference between the bulk thermodynamical pressure $\Pb$ and the
kinetic pressure on the wall $ \kb T \sum_\alpha \roa (x=b)$. As shown for
instance in Ref. \cite{Carn&Chan80}, 
\begin{equation}
\begin{split}
  \label{contact}
  \beta\Pb = & \sum_\alpha \roa (x=b)  
  - 2\pi\dw \beta\left[\int_b^{+\infty} dx
  \, \sum_\alpha \ea\roa (x)\right]^2\\
  & - \beta\int_b^{+\infty} dx
  \, \sum_\alpha \roa (x) \frac{\partial \left[ \ead \Vself 
 \right]}{\partial x}(x) \\
   & - \beta\int_b^{+\infty} dx \int _b^{+\infty} dx' \int d\vy  
  \frac{\partial \left[ \vw - \vb \right] }{\partial x} (x, x',
  \vy)  \\
 & \phantom{\int_b^{+\infty} dx \int _b^{+\infty} dx' \int d\vy}
  \times \sum_{\alpha , \gamma} \ea \eg \roa (x) \rog(x') 
  h_{\alpha \gamma} (x, x', \vy) 
\end{split}
\end{equation}
Since the global charge in the vicinity of a dielectric wall vanishes (see
\eqref{sumrule2}), the
second term in the  r.h.s. of \eqref{contact} is equal to zero. 

The  contact theorem implies that compensations
between the various terms in the r.h.s. of \eqref{contact}  ensure that the 
bulk pressure is independent from $b$ as well as from $\ew$; in other words, the
bulk pressure
is independent from the specific forms of the interactions between particles 
and the
wall, whether the latter  interactions are   purely geometric repulsions or
coulombic couplings.

The above compensations can be checked at  first order in the coupling parameter 
$\epsd$, as shown in Appendix. 
On one hand, up to order $\epsd$, the bulk pressure $\Pb$ is just the 
sum of the ideal-gas pressure plus the Debye correction \cite{Deby&Huck23}
\begin{equation}
  \label{PDebye}
  \beta \Pb = \sum_\alpha \roab - \frac{\kd^3}{24 \pi} + \cO \left
  (\rho \epsd^2  \ln \epsd \right)
\end{equation}
where $\rho$ is the order of magnitude of the 
$\rho_{\alpha}^{\scriptscriptstyle B}$'s.
On the other hand, we calculate the r.h.s. of \eqref{contact} by using the fact
that the  density profiles at first order in $\epsd$  obey the first 
BGY equation \eqref{BGY}. In order to handle  the contributions 
from the screened self-energy in $\roa(x)$  properly, the integrals are performed by using properties similar to
\eqref{fundprop} and derived from \eqref{trotrodeux} and \eqref{trotro}. 
Finally, after compensations of  terms 
involving the dielectric response of the
wall, the ideal-gas pressure arises and the remaining term is reduced to the
difference between the kinetic pressure at the contact with a 
plain hard wall ($\ew = 1$) and the ideal-gas pressure. The latter difference 
involves only the explicit value of the screened self-energy at the contact $x=b$ 
with a plain hard wall. This value is independent from $b$ and gives the term of
order $\epsd$ in  the bulk pressure \eqref{PDebye}.

%%%%%%%%%%%%%%%%%%%%%%%%%%%%%%%%%%%%%%%%%%%%%%%%%%%%%%%%%%%%%%%

\section{Case of a plain hard wall ($\ew = 1$)}

\subsection{Explicit formulas}

In this case the profile density is ruled only by the competition
between Coulomb interactions in the fluid and the geometric deformation of
screening clouds by the impenetrable wall. 
According to \eqref{vtotcmlin} and \eqref{rhoexp}, the expression of the density
profile is reduced to 
\begin{equation}
  \label{titu}
  \roahw (x) = \roab \left\{ 1 - \beta \ead \frac{\kd}{2} \Lhw \left
  ( \kd (x-b) \right) - \beta \ea \Phw (x) \right\}
\end{equation}
In \eqref{titu}, $\Lhw$ can be 
explicitly calculated from \eqref{valueL} 
with the result (see \cite{Janco82I})
\begin{equation}
  \label{cdLhwexp}
  \Lhw (u) = e^{-2u} \left[ \frac{1}{2u} +  \frac{1}{u^2}+\frac{1}{2u^3}
       \right] - \frac{1}{u} K_2 (2u) 
\end{equation}
where $K_2$ is a Bessel function.
According to \eqref{ctu}, $\Mhw$ is reduced to $\Mbhw (x)$. Thus $\Mhw$ is
independent from the species $\gamma$ and, according to \eqref{vtotcmlin},
\begin{equation}
  \label{VHW}
  \Phw (x) = - \frac{ 2 \pi \beta \left( \sum_\gamma \rogb \eg^3
    \right)}{\kd}  
 \Mbhw \left( \kd (x -b) \right)
\end{equation}
where $\Mbhw$ depends only on $\kd (x-b)$ since it is defined in terms of 
$\Lb^{\scriptscriptstyle HW}(\kd x;\kd b)$  through \eqref{exprMb} and 
$\Lb^{\scriptscriptstyle HW}(\kd x;\kd b) =\Lhw ( \kd  (x-b))$ according to
\eqref{decompL}. Its expression at $u = \kd (x-b)$
coincides with \eqref{expMb} when $\dw$ is set to zero, 
\begin{equation}
  \label{std}
  \Mbhw (u) =  \int_1^{+\infty} \, dt \, \left[ \frac{ e^{-2tu} -2t
  e^{-u} }{1-(2t)^2} \right] \frac{1}{\left( t +\sqrt{t^2-1}
  \right)^2}
\end{equation}
The large-distance behaviour  of \eqref{std} reads
\begin{equation}
  \label{}
  \Mbhw (u) \underset{ u \rightarrow \infty}{\sim} \frac{1}{2}
  \left[ \ln 3 - 2 + \frac{\pi}{\sqrt{3}} \right] e^{-u}
\end{equation}
Since $\Mbhw $ is a positive function, $\Phw (x)$ has the same sign at all
distances from the wall. This sign is determined by $\sum_\gamma \rogb \eg^3$,
\begin{equation}
  \label{signphi}
 \left( \sum_\gamma \rogb \eg^3 \right) \Phw (x)<0
\end{equation}

%%%%%%%%%%%%%%%%%%%%%%%%%%%%%%%%%%%%%%%%%%%%%%%%%%%%%%%%%%%%%%%

\subsection{Electric layers}

 Near a hard wall the charge density profile takes the simple form 
\begin{multline}
  \label{chargeHW}
  \sum_\alpha \ea \roahw (x) \\
 = - \beta \left( \sum_\gamma \rogb
 \eg^3 \right) \frac{\kd}{2} \left[ \Lhw \left( \kd (x-b)
  \right) - \Mbhw \left(\kd (x -  b) \right) \right] 
\end{multline}
where $-\Mbhw$ is proportional to $\Phw$ according to \eqref{std}.
We have checked that \eqref{chargeHW} agrees with the result (32) in
Ref. \cite{Guer70}. Since 
$\Lhw(0) = 1/3$ and $\Mbhw (0) \- = \left[  \ln 3 + 1 - \pi / \sqrt{3} \right] 
/8$, 
$\Lhw(0) > \- \Mbhw (0)$, 
while $\Lhw$ and $\Mbhw$ are positive functions of
 $x$ which decay respectively as $\exp (-2 \kd x )/x$ and  $\exp ( - \kd   x)$ 
when $x$ goes to infinity. Therefore the expression \eqref{chargeHW} implies
  that the charge density profile is at least a double layer.
  
In  a charge-symmetric plasma in a symmetric state, the local neutrality
\eqref{nocharge} and  the vanishing of 
$\Phi(x)$ \eqref{noV}, which are valid
whatever the strength of the coupling inside the plasma may be and for
any value of the densities, are retrieved at first
order in $\epsd$ from our expressions. Indeed, according to \eqref{chargeHW} 
and \eqref{VHW},  $\sum_\alpha \ea \roax$ and $\Phw (x)$ are both
proportional to $\sum_\gamma \rogb \eg^3$ and this combination vanishes
in any charge-symmetric plasma in a symmetric state. In a plasma with 
an even or odd number of species, for a particular set of densities 
which satisfies the constraint
\begin{equation}
  \label{const}
  \sum_\gamma \rogb \eg^3 =  0 
\end{equation}
the properties \eqref{nocharge} and \eqref{noV} happens to be valid at
first order in $\epsd$.

In a plasma which is not in a symmetric state, 
the charge density profile does not vanish and
the  sign of the charge density at the wall is fixed by the sign of 
$\sum_\gamma  \rogb \eg^3$ (in the considered weak-coupling regime),
 \begin{equation}
    \label{signun}
    \left( \sum_\gamma \rogb \eg^3 \right) \sum_\alpha \ea \roahw (x=b)
    < 0 
  \end{equation}
The inequality \eqref{signun} implies that if the magnitudes of
positive charges is far larger than those of negative charges, then the
layer at the contact with the wall is negatively charged. 
Moreover, the
combination of \eqref{signphi} and \eqref{signun}
implies that
\begin{equation}
  \label{signdeux}
  \Phw (x=b) \sum_\alpha \ea \roahw (x= b) >0 
\end{equation}

%%%%%%%%%%%%%%%%%%%%%%%%%%%%%%%%%%%%%%%%%%%%%%%%%%%%%%%%%%%%%%

\subsection{Repulsion from the wall for the total particle density}

According to the bulk local neutrality \eqref{localneutr},
\begin{equation}
  \label{kinplain}
  \sum_\alpha \roahw (x) = \sum_\alpha \roab - \frac{\kd^3}{8 \pi} \Lhw
  \left(\kd(x-b) \right) 
\end{equation}
Since $\Lhw$ is a positive function, the total particle density is lower
than its bulk value at any point,
\begin{equation}
  \label{totpart}
  \sum_\alpha \roahw (x) < \sum_\alpha \roab
\end{equation}
The total particle density undergoes
a repulsion at any distance. 

The repulsion from the wall also operates for every particle density when the
electrostatic potential $\Phi(x)$  vanishes, namely in the 
case of a charge-symmetric plasma in a symmetric state or in a plasma where the
composition happens to satisfy \eqref{const}.
Indeed, according to \eqref{titu}, at
any point
\begin{equation}
  \label{}
  \left. \roahw (x) \right|_{\sum_\gamma \rogb \eg^3 =  0} < \roab 
\end{equation}
The repulsion from the wall for every species when the potential drop
with the bulk vanishes is interpreted as follows. According
to \eqref{rocmm}, when $\Phi (x=b) = 0$, the ratio $\roahw (x=b) / \roab$ is
only determined by the screened self-energy which is the difference between 
the values of 
$(\ead /2) [\phi - \vb] (\vr, \vr)$ at the wall and in the
bulk. According to  \eqref{titu}, this difference is equal to 
$(\ead /2)\kd \Lhw(0)$, which is positive.
 Thus, when there is no potential drop between the wall and the bulk, 
 the immersion free-energy in the bulk is lower than its value at 
 the hard wall 
 : the charge surrounded by its screening cloud 
with global charge of opposite sign
is more stable in the bulk than at the plain wall. In other words, for all species Coulomb screening is less efficient when
polarization clouds are deformed by the presence of the hard wall. As an
illustration, we consider a symmetric two-component plasma made of
charges $e$ and $-e$. According to charge-symmetry $\rho_+ (x) = \rho_- (x) $ 
and the profile density is drawn in Fig. \ref{f8a}:a.

%%%%%%%%%%%%%%%%%%%%%%%%%%%%%%%%%%%%%%%%%%%%%%%%%%%%%%%%%%%%

\subsection{Particle density profiles}

The case of plasmas in a charge symmetric state has been discussed in the previous
subsection. 
Here we consider  the generic case where $\sum_\gamma \rogb \eg^3 \neq 0$. Then,
already at leading order in $\epsd$, $\Phw (x) \neq 0$ and 
$\sum_\alpha \ea \roax \neq 0$ so that the electrostatic potential created by the 
electric layer interplays  with the geometric repulsion from the wall.

If $\sum_\gamma \rogb \eg^3 >0$ then  $\Phw (x) < 0$ for all $x$ according to
\eqref{signphi}.
Thus for all negatively charged species $\alpha^-$ at any point, according
 to \eqref{titu}, 
\begin{equation}
  \label{tata}
  \rho_{\alpha^-}^{{\scriptscriptstyle HW}}(x) < \rho_{\alpha^-}^{\ssB}
\end{equation}
because the geometric and electrostatic
effects are both repulsive for them. 
For positively charged particles, since $\Lbhw (x)$ decays faster than 
$\Phw (x)$, the attractive effect of the
potential drop overcomes the wall repulsion at sufficiently large
distances, and $\rho_{\alpha^+}^{{\scriptscriptstyle HW}} (x) - \roab$
becomes positive a priori at least 
at some distance $x_0$ before decaying to zero when $x$
goes to $\infty$,
\begin{equation}
  \label{}
  \rho_{\alpha^+}^{{\scriptscriptstyle HW}} (x) - 
 \rho_{\alpha^+}^{\ssB}
\geq 0 \qquad \mbox{for} \quad x \geq x_0
\end{equation}
The result of the competition between the two effects is
given by 
the density on the wall, which reads
\begin{equation}
  \label{rhocontact}
  \roahw (x=b ) = 
 \roab \left\{ 1  - \frac{1}{6} \beta \kd \left[ \ead - \ea
  \frac{3}{4} \frac{\sum_\gamma \rogb \eg^3}{\sum_{\gamma'}
  \rho_{\gamma'}^\ssB e_{\gamma'}^2} \left( \ln 3 + 1 - \frac{\pi}{\sqrt{3}}
  \right) \right] \right\}
\end{equation}
The sign of $\roahw (x=b) - \roab$ depends on the particular composition
of the plasma.

More precise results about the layer structure of $\roahw (x) - \roab$
can be obtained in two special cases. First, 
in a two-component plasma made of charges $e_+$ and $e_-$, the local
neutrality in the bulk \eqref{localneutr} enforces that 
\begin{multline}
  \label{squ}
  \rohw_+ (x) = \rho_{+}^\ssB \left\{ 1 \rule{0mm}{7mm} \right. 
\left. - \frac{\beta \kd e_+^2}{2}
  \left[ \rule{0mm}{6mm} \Lhw \left( \kd (x-b) \right) \right. \right. \\
\left. \left. - \left[ 1 + \frac{e_-}{e_+}
  \right] \Mhw (\kd (x - b) ) \right] \rule{0mm}{7mm} \right\}
\end{multline}
The argument displayed after \eqref{chargeHW} shows that if $e_+ > |e_-|$, 
$\rohw_+ (x) < \ropb$ 
in a strip $b<x<x_0$ whereas $\rohw_+ (x) > \ropb$ for all $x >x_0$. 
As a conclusion, at leading order in $\epsd$, the wall
repulsion still overcomes the electrostatic attraction arising from
$\Phw$ for the positive charges near the wall and the profile density
$\rop^{{\scriptscriptstyle HW}} (x) - \rho_{+}^\ssB$ has the
structure of a double layer. This can be seen in Fig. \ref{f8b}:b.

At last, we briefly discuss the case of a three-component plasma with
  $\sum_\gamma \eg^3 \rogb > 0$ and which is made for instance of
  species $e_1 > 0 $, $e_2 >0$ and $e_3<0$. Then $\rho_3 (x) <
  \rho_{3}^\ssB$ 
according to \eqref{tata}. However, according to \eqref{rhocontact}, the
composition $\{ \rogb, \eg \}_{\gamma = 1, 2,3}$ of the
  fluid may happen to be such that the electrostatic attraction of
  positive charges (proportional to $\ea$) overcomes the geometric 
repulsion from the wall at  $x=b$ (proportional to $\ead$) 
for the species which carries the positive  charge $e_i$ with the
  lowest magnitude, for instance $e_2$. 
Then $\rho_2(x=b)$ may happen to be larger than $ \rho_{2}^\ssB$
 in spite of the wall geometric repulsion. If 
$\rho_2 (x=b) > \rho_2^{\ssB}$,  
according to \eqref{signun} and the neutrality condition \eqref{localneutr}, 
$ \rho_1 (x=b) < \rho_1^{\ssB}$, and   $\rho_1 (x) - \rho_1^{\ssB}$ 
  contains at least a double layer.

%%%%%%%%%%%%%%%%%%%%%%%%%%%%%%%%%%%%%%%%%%%%%%%%%%%%%%%%%%%%%%%%%

\section{Generic local properties}

\subsection{Large-distance behaviours}

As shown in Section 2 the density profile takes the form \eqref{tita}.
The total screened self-energy $\ead \Vselfsc (x)$ is written in  \eqref{relVL}.
 Its decay at large distances
$\ead \exp ( - 2 \kd x) / 4 x$ (given by \eqref{ch}) is independent from
$\dw$ and is positive~: far away from the wall, the screened self-energy
is a repulsive effect, even if the electrostatic response of the wall
upon one charge is attractive ($\dw > 0$). The contribution from the
complete screened self-energy $\ead \Vselfsc$
to $\roa (x) - \roab$ is drawn in Fig. \ref{fig7}.

The electrostatic potential decays only as 
$\exp (- \kd x) $
  at large distances from the wall,
\begin{equation}
  \label{}
  \Phi (x) \underset{x \rightarrow + \infty}{\sim} \Pas\,
  e^{- \kd x}
\end{equation}
The sign of $\Pas$ depends on the composition 
$\{ \eg, \rogb \}_{\gamma = 1, \ldots, n_s}$ as well as on $\kd b$ and $\dw$. Indeed
\begin{equation}
  \label{}
  \Pas = - \frac{2 \pi \beta }{\kd} \somg \eg^3 \rogb \left\{ \Mb_{{\rm
  as}} ( \kd b, \dw) + \left[ M_\gamma - M \right] ( \xt = 0; \epsg,
  \kd b, \dw) \right\}
\end{equation}
where, in the limit $\eta\equiv \kd b=0$,  $\Mb_{{\rm as}}$ is obtained
from \eqref{expMb}
\begin{multline}
  \label{Mbaras}
  \Mb_{{\rm as}} (\eta=0, \dw) = \frac{1}{8} \frac{1 -
  \dw^2}{1+\dw+\dw^2} \left\{ \frac{\pi}{\sqrt{3}} + (1 + 2 \dw) \ln 3
  \right. \\
\left. + 2 \frac{(1-\dw^2)}{\dw} \ln (1 -\dw) \right\}
\end{multline}
while $\left[ M_\gamma - M \right] ( \xt = 0; \epsg,
  \eta=0, \dw)$ is calculated from
\eqref{decompM}.

In an asymmetric plasma or a charge-symmetric plasma with at least four
components and in an asymmetric state, as far as the density profile of one 
species is concerned, the
$\exp ( - 2 \kd x) / x$ tail of the screened self-energy is always
overcome by the effect of $ - \ea \Phi (x)$, 
\begin{equation}
  \label{}
  \roaasym (x) \equivx \roab \left[ 1 - \beta \ea \Pas  
 e^{- \kd x} \right]
\end{equation}
On the contrary, in a symmetric two-component plasma or in a charge-symmetric
plasma with more than two components and in a symmetric state (see
definition \eqref{defsym}), $\Phi (x) = 0 $ and
\begin{equation}
  \label{}
  \roasym (x) \equivx \roab \left[ 1 - \beta \ead \frac{e^{- 2 \kd x
  }}{4x} \right] 
\end{equation}

However, because of the bulk local neutrality, the influence of 
$\Phi (x)$ at large distances disappears in the total particle density
$\sum_\alpha \roax$ even in the case of an  asymmetric plasma
\begin{equation}
  \label{}
  \soma \roax \equivx \soma \roab - \frac{\kd^2}{16 \pi} \frac{e^{-2 \kd
  x}}{x} + \cO \left( \frac{e^{-3 \kd x}}{x} \right)
\end{equation}
The total particle density is submitted to an effective repulsion far
away from the wall. On the contrary, the charge density at large distances is
determined by $\Phi(x)$,
\begin{equation}
  \label{}
  \soma \ea \roax \equivx - \frac{\kd^2}{4 \pi} \Pas e^{-\kd x} 
\end{equation}
When approaching the bulk region, the charge density
vanishes with a sign ruled by the composition of the Coulomb fluid.

%%%%%%%%%%%%%%%%%%%%%%%%%%%%%%%%%%%%%%%%%%%%%%%%%%%%%%%%%%%%%%%%%%%%

\subsection{Effect of the wall dielectric response}

The three effects which
interplay in the density profiles when $\ew \neq 1$ have
been discussed in  subsections 5.3 and 6.1. Here we consider an asymmetric
two-com\-po\-nent plasma with $e_+ = - 2 e_{-}$ and we comment briefly the
corresponding figures for the profiles of particle  and charge densities.

First we turn to the density profiles. 
When $\ew = 1$ the density at the wall ($x=b$) differs
from its bulk value only by a term of order $\epsd = \kd \beta e^2 / 2$
(see Fig. \ref{f8}:b).
When $\ew <1$ 
the electrostatic repulsion from the wall makes all density profiles
at the wall ($x=b$) vanish exponentially fast when $b$ goes to zero (see
Fig. \ref{f10a}:a).
When $\ew >1$ the density at $x=b$ increases as
$\ew$ gets larger because of the electrostatic attraction to the wall
(see Fig. \ref{f10b}:b). Subsequently, the difference $\rop (x) - \ropb$, which
has a double-layer structure when $\ew \leq 1$, exhibits a
threefold-layer structure when $\ew$ becomes sufficiently large. 

The charge density profile $C(x) = e_+ \rop(x) + e_- \rom (x)$ obeys the
same evolution when $\ew$ varies. If $\ew =1$, according to \eqref{signphi}, the condition $\rop e_+^3
+ \rom e_-^3 >0$ implies that $\Phw (x=b) <0$ and enforces the double
layer $ \ominus \, \oplus $ shown in Fig.\ref{f9} and discussed after
\eqref{chargeHW}. This double layer arises from the balance between the
electrostatic force associated with $\Phi(x)$ and created by $C(x)$
itself and the geometric repulsion from the wall due to the deformation
of screening clouds. When $\ew <1$, the extra electrostatic repulsion
from the wall does not destroy the double layer and only enforces the
vanishing of $C(x)$ at $x=0$ when $b=0$ (see Fig. \ref{f11a}:a). 
When $\ew
>1$, in the case of positive charges, the electrostatic self-attraction
to the wall competes with the opposite effect of $\Phi (x)$ and for high
enough values of $\ew$, $C(x)$ contains three layers $ \oplus \, \ominus 
\, \oplus$ (see Fig. \ref{f11b}:b). When $\ew$ becomes far larger, the 
electrostatic self-attraction to the wall is so strong that the sign 
of $\Phi(x)$ at large distances changes and again there appears a double
layer $ \oplus \, \ominus$ but with signs reversed with respect to the 
situation when $\ew = 1$ (see Fig. \ref{f11c}:c). 

%%%%%%%%%%%%%%%%%%%%%%%%%%%%%%%%%%%%%%%%%%%%%%%%%%%%%%%%%%%%%%%%%%%%%%%%%%%%%%%%
\appendix

\section{Appendix }

In the present appendix we show that the contact theorem \eqref{contact} 
is satisfied by the
profile densities  found at first
order in $\epsd$. First, we consider a plain hard wall ($\dw=0$) and then we
turn to the case of a wall with a dielectric response.

In the case of a plain hard wall, the r.h.s. side of \eqref{contact} is reduced
to the kinetic pressure on the wall
\begin{equation}
  \label{contplain}
\beta\Pb
= \sum_\alpha \left.\roa\right\vert_{\dw=0} (x=b)  
\end{equation}
According to the bulk local neutrality
\eqref{localneutr} the contribution from $\Phi(b)$ to
$\sum_\alpha \left.\roa\right\vert_{\dw=0} (x=b)$ is zero, and, 
according to
\eqref{kinplain},
\begin{multline}
  \label{Pcinplain}
\sum_\alpha \left.\roa\right\vert_{\dw=0} (x=b)
=\sum_\alpha \roab 
- \beta \sum_\alpha \roab \ead
\left. \Vselfsc\right\vert_{\dw=0}(x=b) \\
+ \cO \left
  ( \rho\epsd^2  \ln \epsd \right)   
\end{multline}
Since $\left. \Vselfsc\right\vert_{\dw=0}(x=b) =-\kd/6$ (see Section 5), 
the kinetic pressure \eqref{Pcinplain} on the wall
does coincide with the value \eqref{PDebye} of the bulk pressure at first order
in $\epsd$.

Now we turn to the case of a wall with a dielectric response. 
First, the last integral in the r.h.s. side of \eqref{contact} can
be written as the sum 
$I_{\rm \scriptscriptstyle W}-I_{\rm \scriptscriptstyle B}$ where 
\begin{equation}
  \label{defIw}
I_{\rm \scriptscriptstyle W}
= \int_b^{+\infty}dx
\sum_{\alpha }  \roa (x) 
J_{{\rm \scriptscriptstyle W},{\alpha}}(x) 
\end{equation}
and $J_{{\rm \scriptscriptstyle W},{\alpha}}(x)$ is the integral in the r.h.s. of the first 
BGY equation \eqref{BGY}. $I_{\rm \scriptscriptstyle B}$ 
and $J_{{\rm \scriptscriptstyle B},{\alpha}}(x)$ are defined in a similar way. 
The value  
$J_{{\rm \scriptscriptstyle W},{\alpha}}^{\scriptscriptstyle MF}(x)$
of 
$J_{{\rm \scriptscriptstyle W},{\alpha}}(x)$ in the mean-field approximation
is given in \eqref{coucouu}. By use of the definitions 
\eqref{defVself} and \eqref{fcmmvb}, it can be rewritten as
\begin{equation}
  \label{valueJw}
J_{{\rm \scriptscriptstyle W},{\alpha}}^{\scriptscriptstyle MF}(x)
=-\beta \ead\frac{\partial}{\partial x} \left[ \Vselfsc (x)- \Vself(x)
    \right] 
\end{equation}
Therefore
\begin{multline}
  \label{sumBIw}
I_{\rm \scriptscriptstyle W}^{\scriptscriptstyle MF}
- \beta\int_b^{+\infty} dx
  \, \sum_\alpha \roa (x) \frac{\partial \left[ \ead \Vself 
 \right]}{\partial x}(x)\\
 =-\beta\int_b^{+\infty} dx
  \, \sum_\alpha \roa (x) \frac{\partial \left[ \ead \Vselfsc 
 \right]}{\partial x}(x)
\end{multline}
The r.h.s. of equality \eqref{sumBIw} does not involve the bare self-energy 
$\Vself(x)$ and it is calculated by 
inserting the expression 
\eqref{tita} of 
$\roa (x)$ in terms of the screened self-energy and the electrostatic potential 
$\Phi(x)$. After an integration by parts, we notice that, according to 
\eqref{trotrodeux} and \eqref{trotro},
\begin{equation}
  \label{trotrotrois}
\int_b^{+\infty} dx \exp[-\beta \ead \Vselfsc(x)]
\frac{\partial \Phi(x)}{\partial x}
=\left[\int_b^{+\infty} dx \frac{\partial \Phi(x)}{\partial x}
\right]\times \left[1+\cO \left( \epsd \ln \epsd \right)\right]
\end{equation}
Then the local neutrality in the bulk \eqref{localneutr} implies that the sum 
\eqref{sumBIw} is equal to
\begin{equation}
  \label{valueBIw}
\sum_\alpha \roab 
- \sum_\alpha \roa (x=b)
+ \cO \left
  ( \rho\epsd^2 \ln \epsd \right)    
\end{equation}

Second, we calculate  $I_{\rm \scriptscriptstyle B}$ which is defined 
in
terms of 
\begin{equation}
  \label{defJB}
 J_{{\rm \scriptscriptstyle B},{\alpha}}\equiv
 - \beta e_\alpha \int_b^{+\infty}dx'\int d\vy
 \, \frac{\partial \vb}{\partial x} ( x-x',\vy) 
 \sum_\gamma e_\gamma \rho_\gamma (x') h_{\alpha \gamma} (x, x', \vy) 
\end{equation}
as in \eqref{defIw}. In fact, 
the difference 
\begin{equation}
  \label{difh}
h_{\alpha \gamma} (x, x', \vy)-
\left. h_{\alpha \gamma}\right\vert_{\dw=0}  (x, x', \vy) 
\end{equation}
gives a vanishing  contribution  to $I_{\rm \scriptscriptstyle B}$. Indeed, the
mean-field value \eqref{MF} of $h_{\alpha \gamma}$ is proportional to
$\fez$ and, according to \eqref{defphi10} and \eqref{valuehop},
$\fez-\left.\fez\right\vert_{\dw=0}$  is bounded by a product of functions
$f(x,x')g(\vy)$ where $g(\vy)$ is integrable while $f(x,x')$ decays exponentially fast in all
directions  in the plane of
variables $(x,x')$. As a consequence 
\begin{equation}
\frac{\partial  \vb  }{\partial x} (x-x', \vy) 
  \sum_{\alpha , \gamma} \ead \egd \roa (x) \rog(x') 
  \left[\fez (x, x', \vy) -
  \left.\fez \right\vert_{\dw=0}(x, x', \vy)\right]
\end{equation}
is absolutely integrable in the space
$(x,x',\vy)$. Moreover, $\fez-\left.\fez\right\vert_{\dw=0}$ 
is symmetric under exchange of $x$ and $x'$
whereas $\partial \vb/\partial x$ is antisymmetric under the same exchange.
Subsequently,  the contribution from the difference \eqref{difh} to 
$I_{\rm \scriptscriptstyle B}$ is just zero. Moreover, when $\dw=0$ 
the self-energy is bounded for all $x$'s ranging from $0$ to $+\infty$ so that,
according to \eqref{vtotcmlin},
 
\begin{equation}
\roa(x)-\left.\roa\right\vert_{\dw=0}(x)=
\roab \left\{\exp\left[-\beta\ea \Vselfsc(x)\right]-1
+ {\mathcal O}_{{\rm exp}} \left( \epsd  \right)\right\} 
\end{equation}
and, according to 
\eqref{trotrodeux} and \eqref{trotro},
$\roa(x)-\left.\roa\right\vert_{\dw=0}(x)$ 
contributes to 
$I_{\rm \scriptscriptstyle B}$  by a term of order
$\rho\epsd^2  \ln \epsd$, as well as
$\rog(x')-\left.\rog\right\vert_{\dw=0}(x')$.
Eventually,
\begin{equation}
I_{\rm \scriptscriptstyle B}=
\left.I_{\rm \scriptscriptstyle B}\right\vert_{\dw=0}
+ \cO \left( \rho\epsd^2  \ln \epsd \right)
\end{equation}
Since the Ursell function
$\left. h_{\alpha \gamma} \right\vert_{\dw=0}$ obeys the BGY hierarchy with
$\vw$ replaced by $\vb$, a calculation similar to that performed for 
$I_{\rm \scriptscriptstyle W}$ leads to
\begin{equation}
  \label{valueIb}
\left.I_{\rm \scriptscriptstyle B}\right\vert_{\dw=0}
=\sum_\alpha \roab 
- \sum_\alpha \left.\roa\right\vert_{\dw=0} (x=b)
+ \cO \left
  ( \rho\epsd^2 \ln \epsd \right)   
\end{equation}
According to \eqref{Pcinplain}, the r.h.s. of \eqref{valueIb} coincides with the
opposite of the term of order $\epsd$ in the bulk pressure. Therefore 
\eqref{valueIb} leads to
\begin{equation}
  \label{valueIbter}
I_{\rm \scriptscriptstyle B}
= \frac{\kd^3}{24\pi }
+ \cO \left
  ( \rho\epsd^2  \ln \epsd \right)  
\end{equation}
Eventually, the sum of the terms in the r.h.s. of \eqref{contact} does coincide
with the value \eqref{PDebye} of $\beta\Pb$.

%%%%%%%%%%%%%%%%%%%%%%%%%%%%%%%%%%%%%%%%%%%%%%%%%%%%%%%%%%%%%%%%%%%%%%%%%%%%

%profil potentiel
\begin{figure}
\centering\epsfig{figure=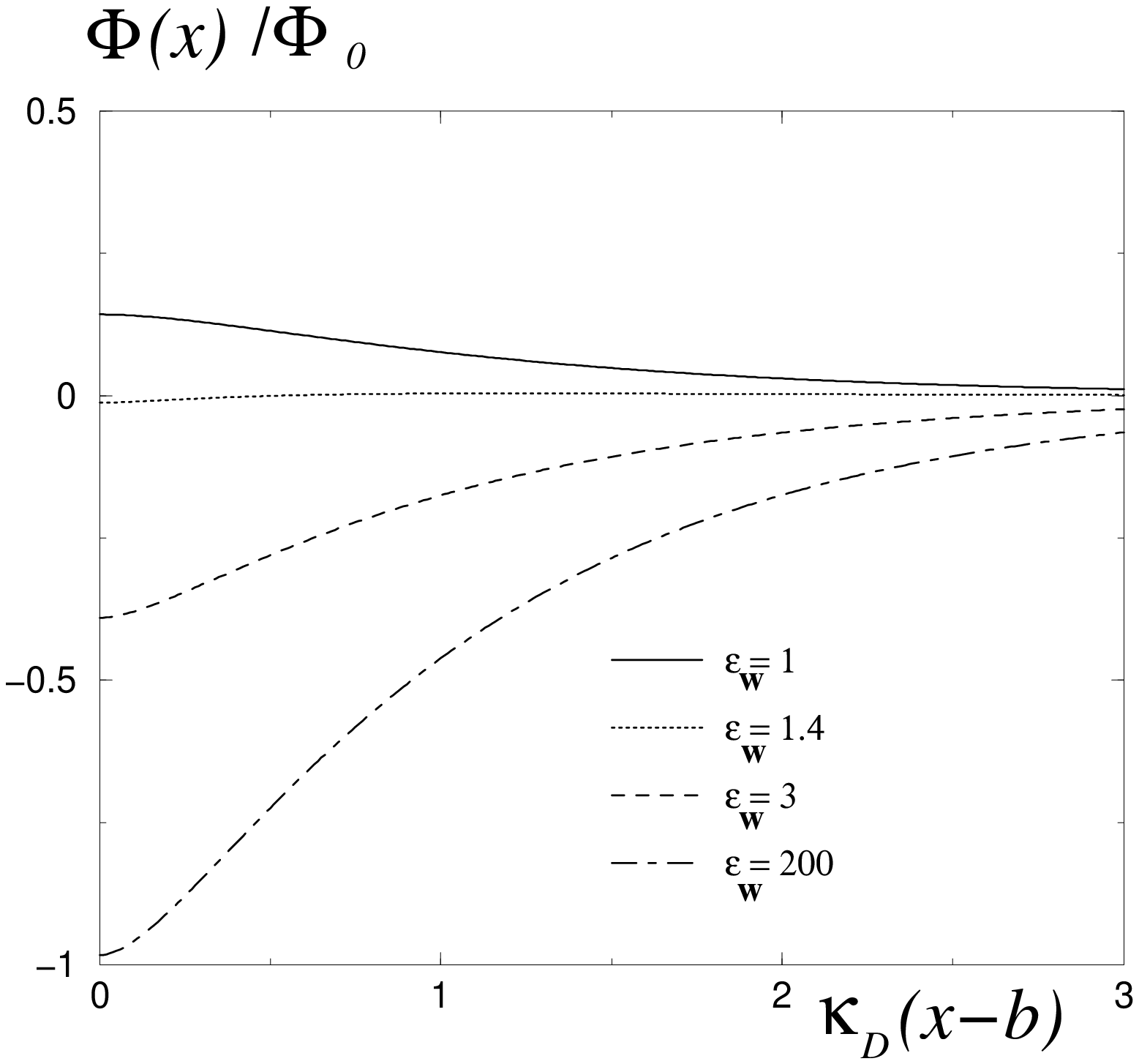, width=6cm}
\caption{Profile of the electrostatic potential $\Phi(x)$ for $\kd b = 0.1$ 
in the limit
  $\dw \beta e^2 / b \ll 1$.  $M_\gamma$ in \eqref{decompM} does 
  no longer depend on $\gamma$ in the limit $\dw \beta e^2 / b \ll 1$. 
  $\Phi_0 = -
  \somg \beta \eg^3 \rogb / \kd $.   
The values of $\ew$ are displayed in the figure.}\label{f6}
\end{figure}

%hard wall

\begin{figure}
\begin{minipage}[b]{0.46\linewidth}
\centering\epsfig{figure=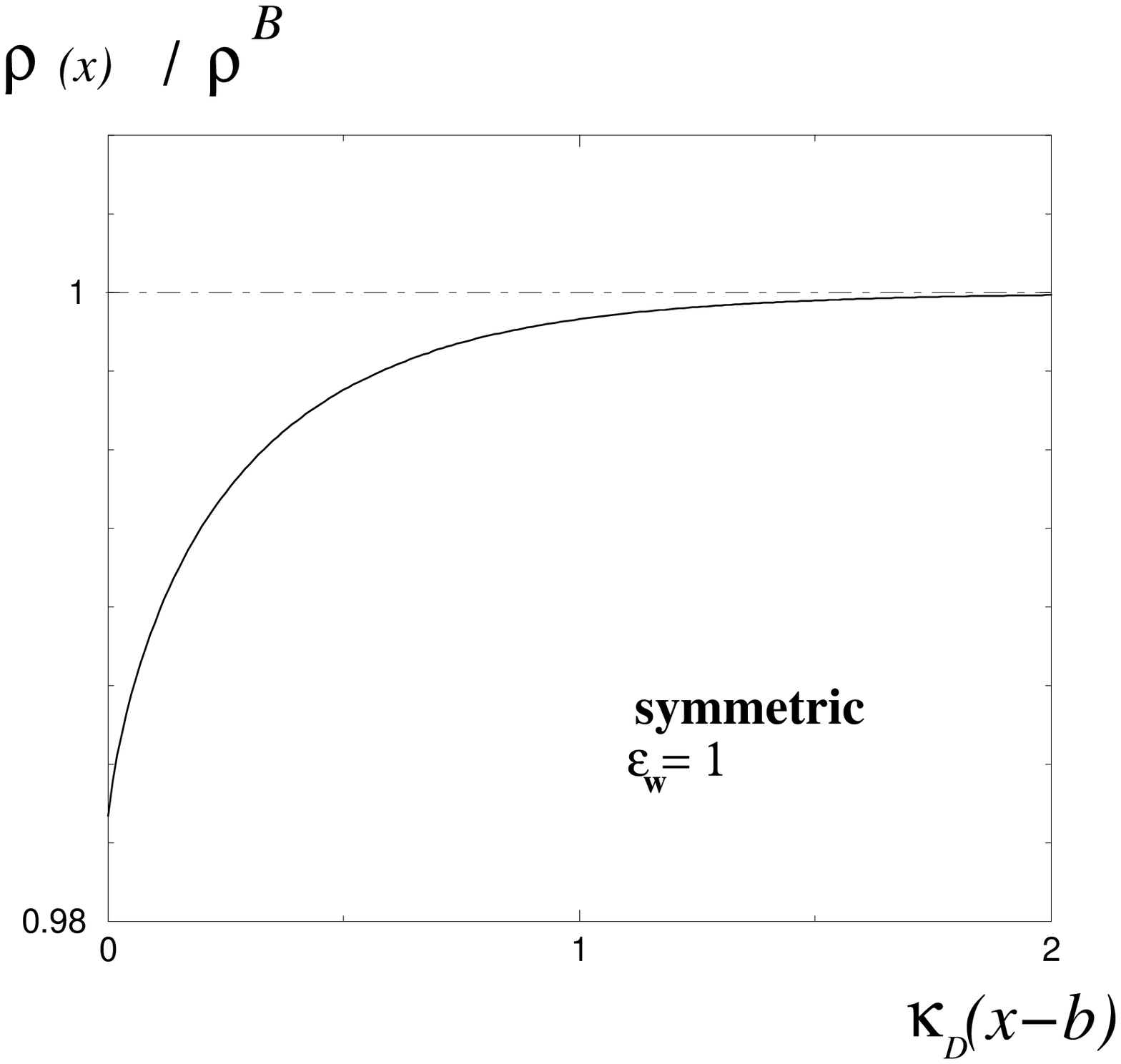, width=5cm}
\caption{a}\label{f8a}
\end{minipage}\hfill
\begin{minipage}[b]{0.46\linewidth}
\centering\epsfig{figure=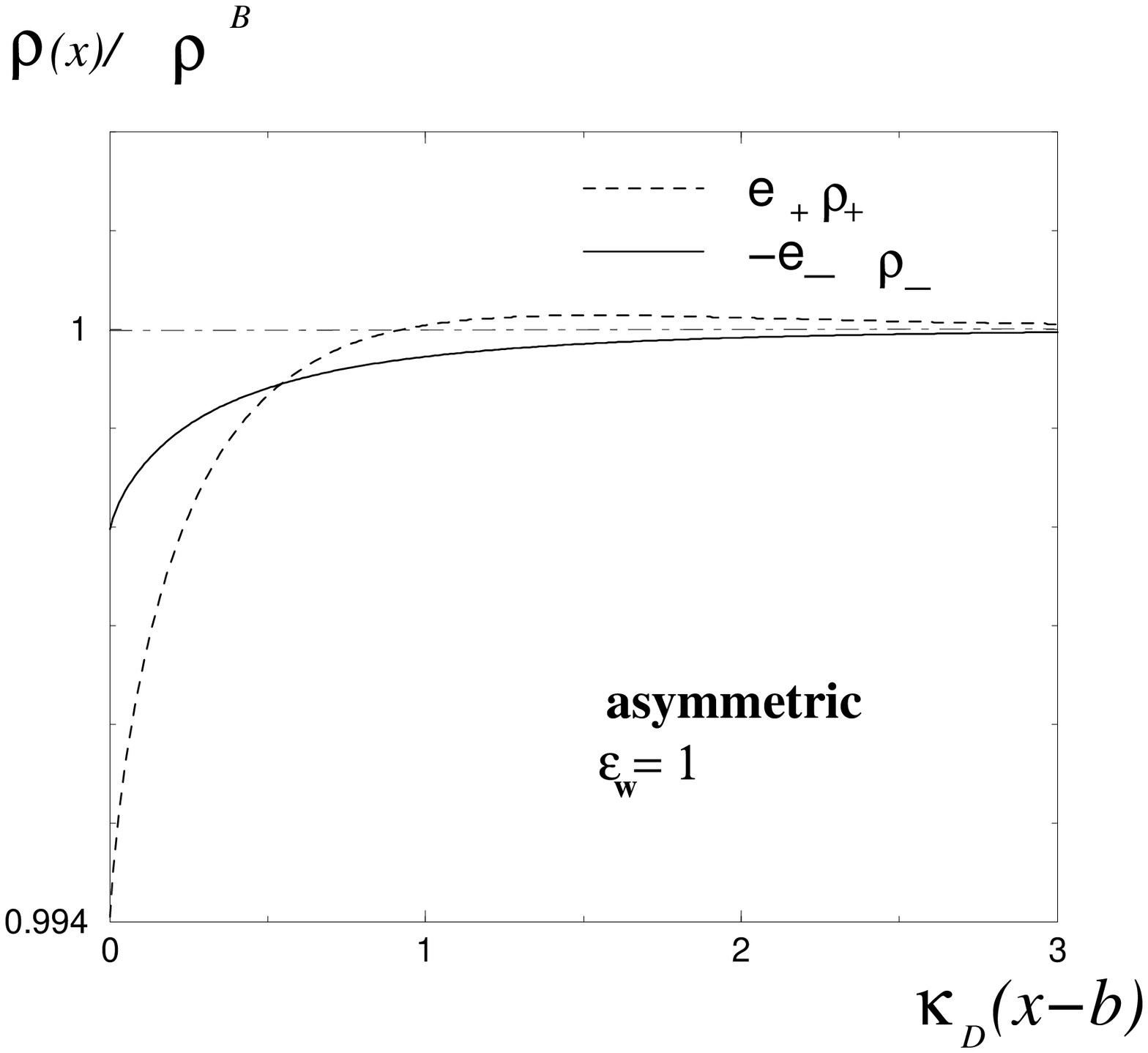, width=5cm}
\addtocounter{figure}{-1}
\caption{b}\label{f8b}
\end{minipage}
\addtocounter{figure}{-1}
\caption{Density profiles in a two-component plasma near a plain hard
  wall. If the plasma is symmetric (Fig. \ref{f8a}:a) $\Phw (x) = 0$ and
  only the geometric repulsion from the wall is involved. The curve is
  the same as in Ref.\cite{Janco82II}. If the plasma is charge
  asymmetric with $e_+ = 2 |e_-|$ (Fig. \ref{f8b}:b), 
 $\Phw (x=b) < 0 = \Phw
  (x = +\infty)$ and the competition between the geometric repulsion
  from the wall and the attraction to the wall by $\Phw (x)$ for
  positive charges results into a double layer structure for $\rop(x) -
  \ropb$. In Figs \ref{f8a}:a and \ref{f8b}:b 
 $\kbe = 0.01$ and $\kdbf=0.1$.}
\label{f8}
\end{figure}

%selfenergy

\begin{figure}[ht]
\begin{minipage}[b]{0.46\linewidth}
\centering\epsfig{figure=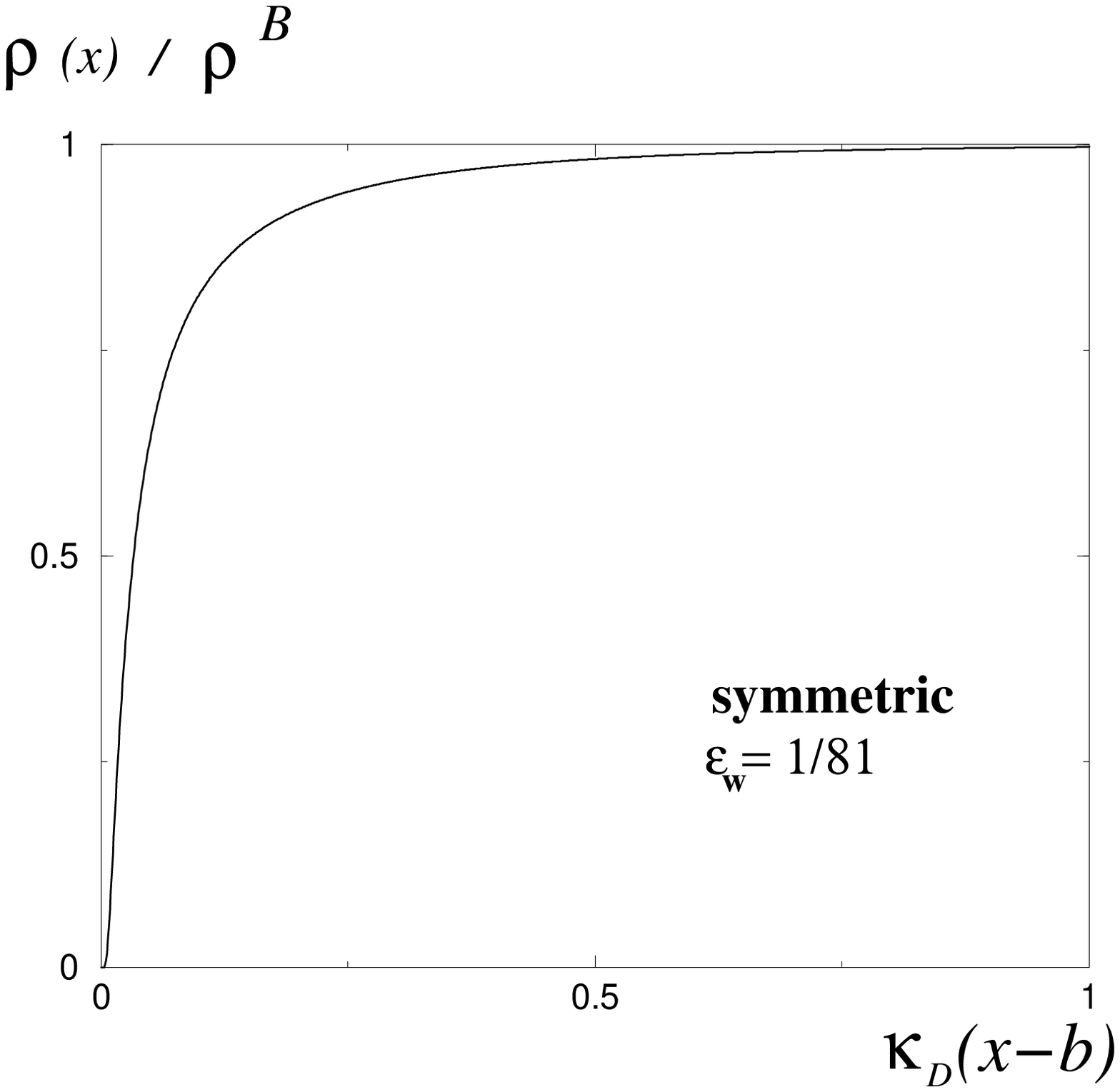,width=5cm}
\caption{a}\label{fig7a}
\end{minipage}\hfill
\begin{minipage}[b]{0.46\linewidth}
\centering\epsfig{figure=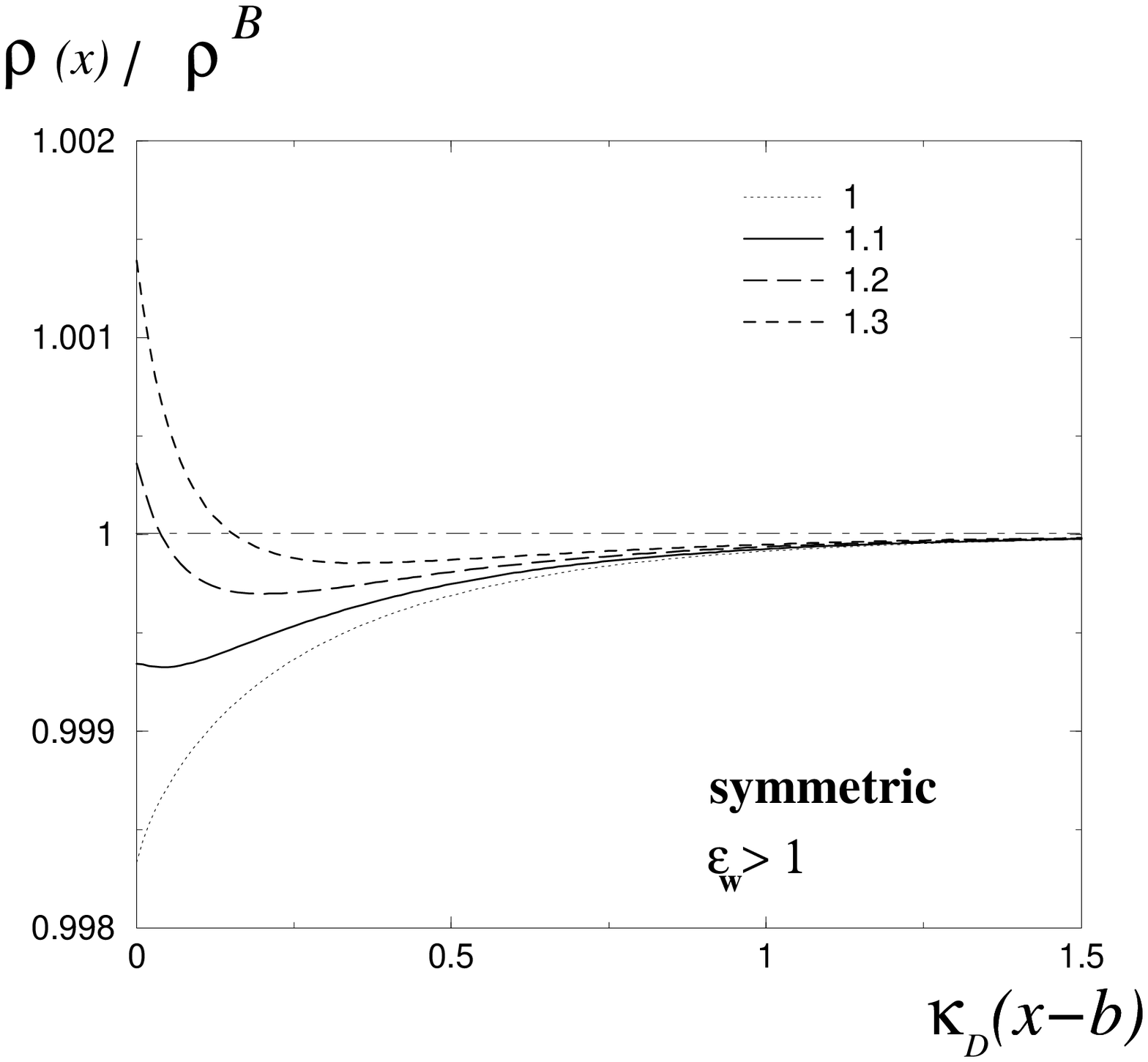, width=5cm}
\addtocounter{figure}{-1}
\caption{b}\label{fig7b}
\end{minipage}
\addtocounter{figure}{-1}
\caption{Contribution from the complete screened self-energy 
$\ead  \Vself$ to the profile density. In the case of a symmetric
  two-component plasma of charges $e$ and $-e$, 
$\rop (x)=\rom (x)=\rho(x)$ and $\Phi (x)=0$ while 
$\rho(x)=\rho^\ssB \exp \left[-\beta e^2 \Vself^{{\scriptscriptstyle sc}}(x)\right]$.
In Fig.  \ref{fig7a}:a,
  where $\ew < 1$, both the electrostatic and geometric repulsions from
  the wall make $\rho(x) < \rho^\ssB $. Fig.  \ref{fig7b}:b, 
 where $\ew >1$, 
  displays the competition at short distances
between the electrostatic attraction to the
  wall (which gets larger when $\ew$ increases) and the geometric
  repulsion from the wall. In Fig. \ref{fig7a}:a, $\kbe = 0.1$ and 
 $\kdbf = 10^{-5}$ whereas 
 in Fig.  \ref{fig7b}:b, $\kbe = 0.01$, $\kdbf = 0.1$ and the values of
  $\ew$ are written in the figure.}
\label{fig7}
\end{figure}

%profil de densite

\begin{figure}
\begin{minipage}[b]{0.46\linewidth}
\centering\epsfig{figure=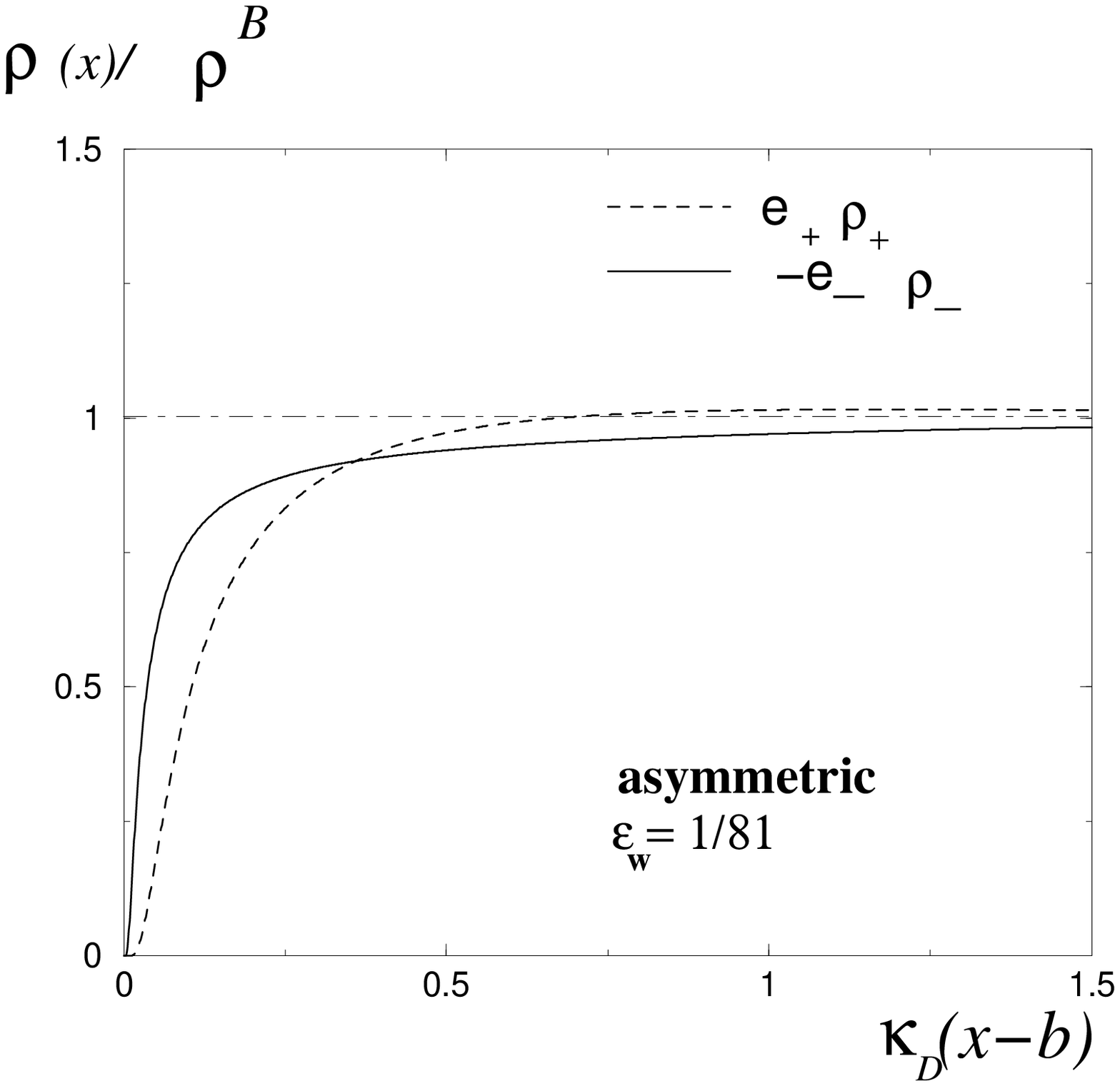, width=5cm}
\caption{a} \label{f10a}
\end{minipage}\hfill
\begin{minipage}[b]{0.46\linewidth}
\centering\epsfig{figure=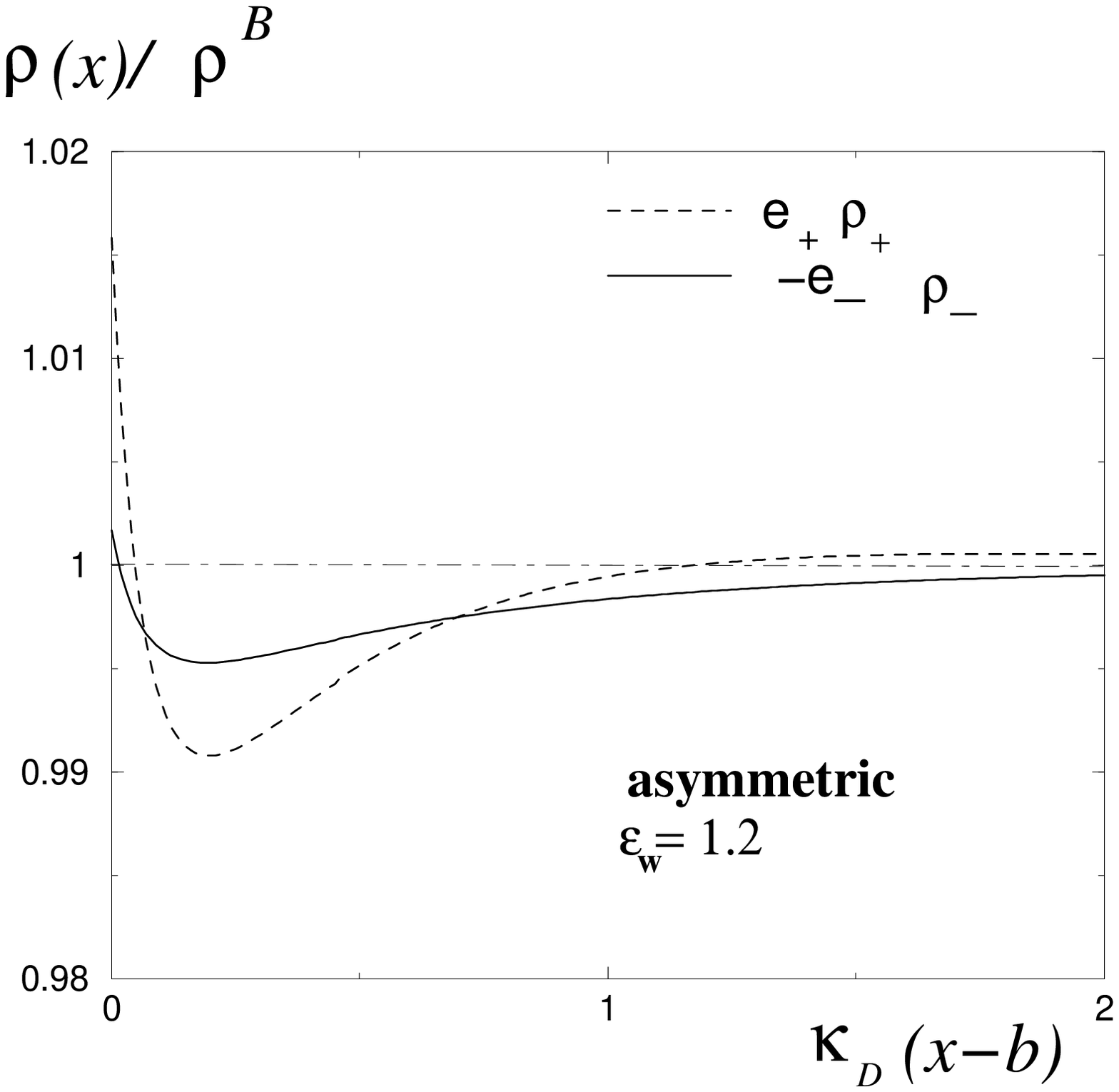, width=5cm}
\addtocounter{figure}{-1}
\caption{b}\label{f10b}
\end{minipage}
\addtocounter{figure}{-1}
\caption{Density profiles in an asymmetric two-component plasma ($e_+ = 2
  |e_-|$) when $\ew$ varies. In Fig. \ref{f10a}:a,  
$\kbe = 0.1$ and $\kdbf = 10^{-5}$, whereas in Fig.  \ref{f10b}:b, 
$\kbe = 0.01$, $\kdbf = 0.1$ and the values of $\ew$ are given
  in the figure.  }
\label{f10}
\end{figure}

%profil de densite de cahrge

\begin{figure}
\begin{minipage}[b]{0.96\linewidth}
\centering\epsfig{figure=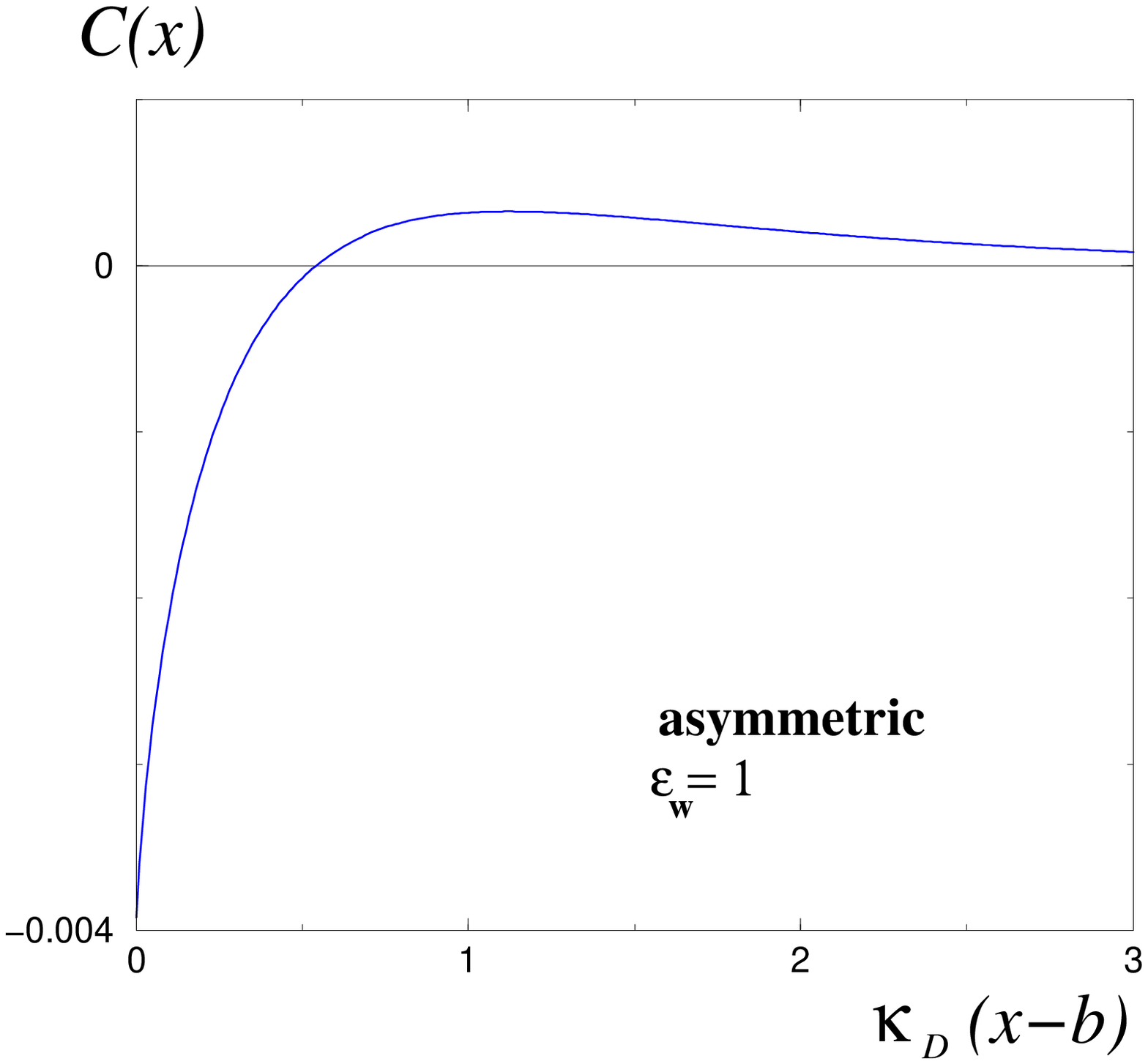, width=6cm}
\caption{Double layer of the charge density profile $C(x) = e_+ \rop (x)
  + e_- \rom (x)$ for the asymmetric two-component plasma already
  considered in Fig. \ref{f8}. $\sum_\gamma \rogb \eg^3 >0$ in this case and
  inequality \eqref{signun} can be checked.}\label{f9}
\end{minipage}
\end{figure}
%profil de densite de charge suite

\begin{figure}
\begin{minipage}[b]{0.96\linewidth}
\centering\epsfig{figure=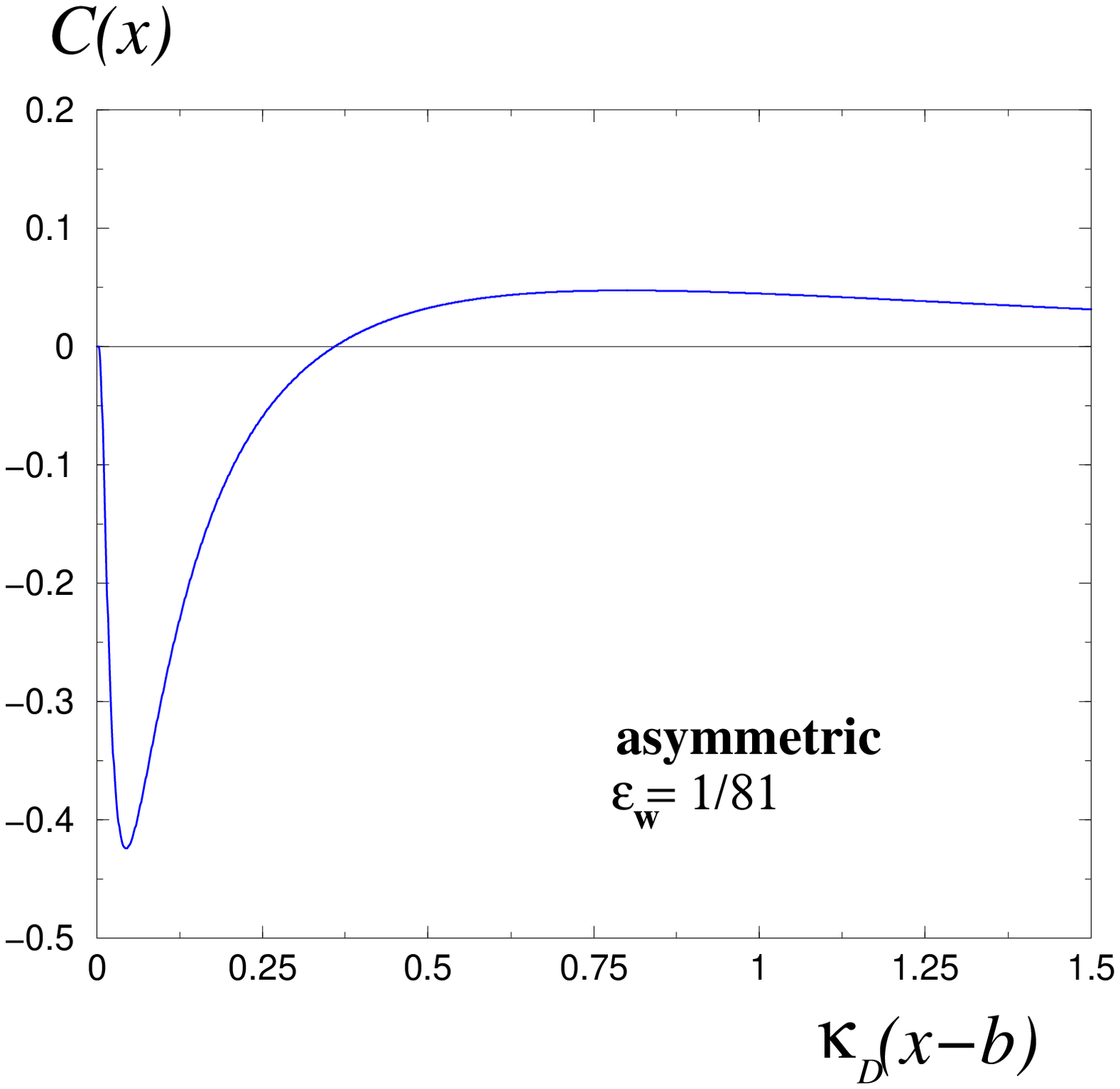, width=6cm}
\caption{a}\label{f11a}
\end{minipage}\\
\begin{minipage}[b]{0.46\linewidth}
\centering\epsfig{figure=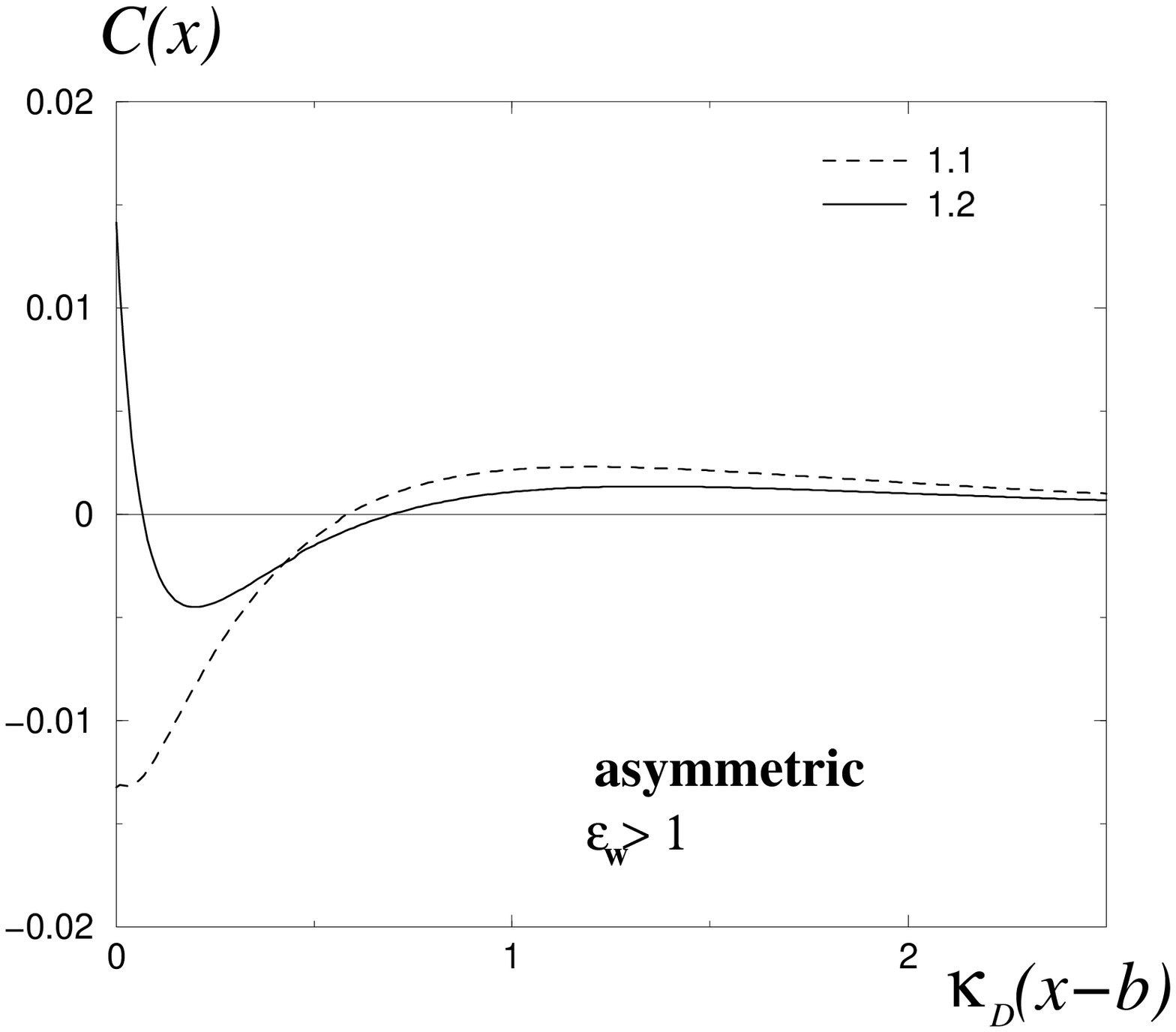, width=5cm}
\addtocounter{figure}{-1}
\caption{b}\label{f11b}
\end{minipage}\hfill
\begin{minipage}[b]{0.46\linewidth}
\centering\epsfig{figure=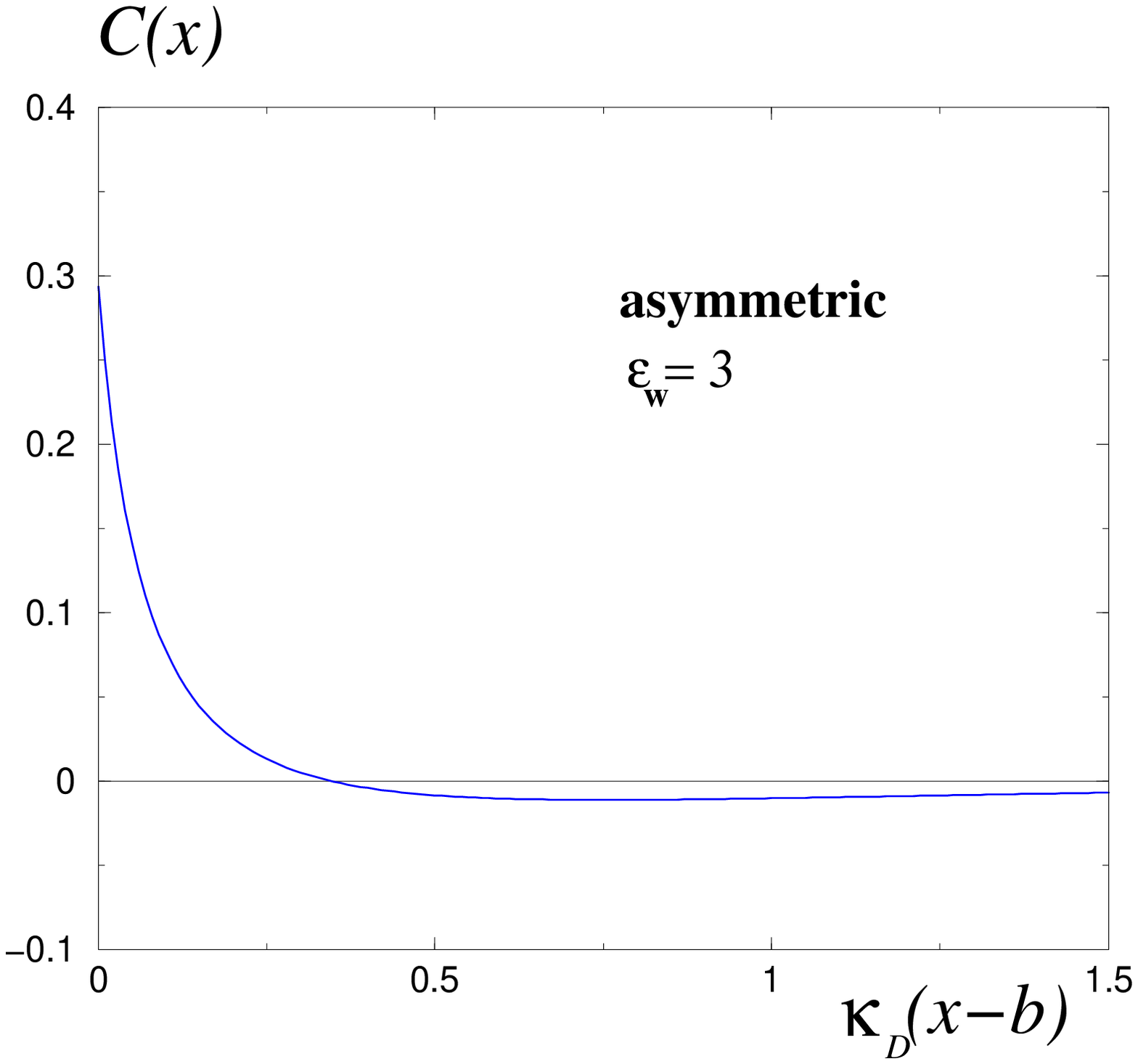, width=5cm}
\addtocounter{figure}{-1}
\caption{c}\label{f11c}
\end{minipage}
\addtocounter{figure}{-1}
\caption{Charge density profiles in the same two-component plasma as in
  Fig \ref{f9}. In  Fig. \ref{f11a}:a , $\kbe = 0.1$ and $\kdbf =
  10^{-5}$ and in Figs. \ref{f11b}:b and \ref{f11c}:c, $\kbe = \kdbf =
  0.1$. }
\label{f11}
\end{figure}

%%%%%%%%%%%%%%%%%%%%%%%%%%%%%%%%%%%%%%%%%%%%%%%%%%%%%%%%%%%%%%%%%%%%%%%%%%%

\bibliographystyle{plain}
\bibliography{densityI}

\begin{thebibliography}{10}

\bibitem{Alas83}
A.~Alastuey.
\newblock The one component plasma near a hard wall: Weak coupling limit with
  image forces.
\newblock {\em Mol. Phys}, {\bf 50}:33, 1983.

\bibitem{Alas&Pere96}
A.~Alastuey and A.~Perez.
\newblock Virial expansions for quantum plasmas: Fermi-bose statistics.
\newblock {\em Phys. Rev. E}, {\bf 53}:5714, 1996.

\bibitem{Carn&Chan80}
S.~Carnie and D.~Chan.
\newblock The statistical mechanics of the electrical double layer~: {S}tress
  tensor and contact conditions.
\newblock {\em J. Chem. Phys.}, {\bf 74}:1293, 1981.

\bibitem{Cornu98II}
F.~Cornu.
\newblock Quantum plasmas with or without a uniform field. {II.} {E}xact
  low-density free energy.
\newblock {\em Phys. {R}ev. {E}}, {\bf 58}:5293, 1998.

\bibitem{Deby&Huck23}
P.~Debye and E.~Hückel.
\newblock Zur {T}heory der {E}lektrolyte.
\newblock {\em Physik. Z.}, {\bf 9}:185, 1923.

\bibitem{Guer70}
R.L. Guernsey.
\newblock Correlation effects in semi-infinite plasmas.
\newblock {\em Phys. Fluids}, {\bf 13}:2089, 1970.

\bibitem{Jackson}
J.D. Jackson.
\newblock {\em Classical Electrodynamics}.
\newblock Wiley, {N}ew {Y}ork, 1962.

\bibitem{Janco82I}
B.~Jancovici.
\newblock Classical {C}oulomb systems near a plane wall. {I}.
\newblock {\em J. Stat. Phys.}, {\bf 28}:43, 1982.

\bibitem{Janco82II}
B.~Jancovici.
\newblock Classical coulomb systems near a plane wall. {II}.
\newblock {\em J. Stat. Phys.}, {\bf 29}:263, 1982.

\bibitem{Mart88}
Ph. Martin.
\newblock Sum rules in charges fluids.
\newblock {\em Review of {M}odern {P}hysics}, {\bf 60}:1075, 1988.

\bibitem{Onsa&Sama34}
L.~Onsager and N.T. Samaras.
\newblock The surface tension of {D}ebye-hückel electrolytes.
\newblock {\em J. {C}hem. {P}hys.}, {\bf 2}:528--536, 1934.

\end{thebibliography}
 
\end{document}